\documentclass[aps,prd,10pt,twocolumn,nofootinbib]{revtex4}

\usepackage{epsfig}
\usepackage{bm}
\usepackage{latexsym}
\usepackage{natbib}
\usepackage{url}
\usepackage{dcolumn}
\usepackage{xcolor}
\usepackage{amsfonts,amssymb,amsmath}
\usepackage{graphicx,epsfig}
\usepackage{psfrag}
\usepackage{subfigure}
\usepackage{hyperref}
\usepackage{ulem}
\hypersetup{
    bookmarks=true,         
    unicode=false,          
    pdftoolbar=true,        
    pdfmenubar=true,        
    pdffitwindow=false,     
    pdfstartview={FitH},    
    pdftitle={My title},    
    pdfauthor={Author},     
    pdfsubject={Subject},   
    pdfcreator={Creator},   
    pdfproducer={Producer}, 
    pdfkeywords={keyword1, key2, key3}, 
    pdfnewwindow=true,      
    colorlinks=true,       
    linkcolor=red,          
    citecolor=blue,        
    filecolor=magenta,      
    urlcolor=cyan           
}

\begin{document}

\title{Enhancement of axion decay constants in type IIA theory?}

\author{Gaurav Goswami\footnote{gaurav.goswami@ahduni.edu.in}}

\affiliation{School of Engineering and Applied Science, Ahmedabad University, Ahmedabad 380009, India}

\begin{abstract}
We investigate the possibility of enhancement of effective axion decay constant in well controlled constructions in string theory. To this end, we study the dynamics of axions arising in the compactifications of type IIA string theory on Calabi-Yau orientifolds with background fluxes (with non-perturbative effects included to ensure stabilization of all moduli). 
In this setup, we attempt to obtain large effective axion decay constant in two different ways: by searching for a direction in field space in which the potential is sufficiently flat and by arriving at a very explicit stringy embedding of the Kim-Nilles-Peloso (KNP) alignment mechanism. We do not find super-Planckian effective decay constants by either of the approaches. Furthermore, we find that the alignment angle of KNP mechanism can not be made arbitrarily small by adjusting the fluxes.
%
%
%

\end{abstract}

\maketitle

\section{Introduction}
Recently, there have been a number of speculations regarding the possible constraints which any self-consistent theory of Quantum Gravity imposes on the low energy effective theory describing the Universe at longer distances. This has lead to various ideas such as the Weak Gravity Conjecture \cite{ArkaniHamed:2006dz} (and its various manifestations), Swampland conjecture \cite{Vafa:2005ui,Ooguri:2006in} (and refined swampland distance conjecture \cite{Klaewer:2016kiy}), de-Sitter swampland conjecture \cite{Obied:2018sgi} and its refinements (see e.g. \cite{Garg:2018reu,Ooguri:2018wrx} for early papers and \cite{Palti:2019pca} for a recent review). 

Some of the arguments on which these conjectures rest are based on our general expectations from any self consistent theory of Quantum Gravity while others are based on diligent studies of well known string compactifications \cite{Font:2005td,Ibanez:2012zz,Douglas:2006es,Baumann:2014nda} at leading order in $\alpha'$ and $g_s$ with well-controlled sub-leading corrections (see e.g. \cite{Danielsson:2018ztv}, \cite{Cicoli:2018kdo} and \cite{Kachru:2018aqn}).  In many such studies, an important role has been played by compactifications of type IIA theory mostly because in such a setting, all geometric moduli get fixed at the classical level (in this context, see e.g. \cite{Grimm:2004ua,DeWolfe:2005uu,Palti:2008mg} for some early papers on moduli stablization in type IIA  theory, \cite{Hertzberg:2007ke,Hertzberg:2007wc,Caviezel:2008tf,Flauger:2008ad,Blaback:2013fca,Blaback:2018hdo} for some attempts to obtain cosmologically interesting solutions and \cite{Acharya:2006ne,Banks:2006hg,Saracco:2012wc,McOrist:2012yc,Blaback:2010sj,Gautason:2015tig,Andriot:2018ept,Junghans:2018gdb,Banlaki:2018ayh,Cordova:2018dbb,Escobar:2018rna} for some possible concerns about the validity of these solutions). 

Another related subject, which has received a lot of attention is the possible non-existence of super-Planckian axion decay constants in well-controlled regimes of string theory \cite{Banks:2003sx}. This subject is of paramount importance given the possible connection of this to large field cosmic inflation: the only version of inflationary scenarios which can be observationally tested in the foreseeable future.
One must recall that at the level of field theory, one could imagine mechanisms which could boost the axion decay constant to super-Planckian values (such as in \cite{Kim:2004rp}, see also \cite{Choi:2014rja}), the same has not been convincingly established in well-controlled regimes of string theory (see e.g. \cite{Blumenhagen:2017cxt} for a recent work and references). Moreover, it is worth noting that many recent attempts to search large axion decay constants \cite{Palti:2015xra} and large field excursions \cite{Baume:2016psm} in well understood regimes of type IIA theory have motivated the refined swampland distance conjecture \cite{Klaewer:2016kiy}. 
All this is consistent with the literature on Weak Gravity Conjecture \cite{ArkaniHamed:2006dz} which mentions what is referred to as the axionic version of the Weak Gravity Conjecture. 

Keeping this interesting literature in mind, we will attempt to come up with different ways of obtaining large (potentially super-Planckian) effective decay constants in the context of well studied type IIA flux vacua \cite{Palti:2015xra,Baume:2016psm} (see also \cite{Hebecker:2018fln} for a recent work).
First, we will  try to do this by finding directions in axion field space such that the scalar potential along the direction is sufficiently flat by making sure that in such directions, the effective decay constant due to one of the axions is large and the vevs of the saxions corresponding to the rest of axions are so large that their contributions to the scalar potential are negligible. We will find that this can always be done for any fixed choice of fluxes but eventually, we shall argue that this approach for finding enhanced effective decay constant is not going to give desired results. 

There have been many attempts in the past to embedd the well known Kim-Nilles-Peloso mechanism \cite{Kim:2004rp} (also called alignment mechanism \cite{Choi:2014rja}) in string constructions (see e.g. \cite{Long:2014dta,Gao:2014uha} and \cite{Hebecker:2015rya,Palti:2015xra,Hebecker:2018fln}).
In the context of IIA theory, this was recently tried in ref \cite{Palti:2015xra}using some of the ideas first presented in \cite{Hebecker:2015rya}. 
We will use the idea presented in \cite{Palti:2015xra} and \cite{Hebecker:2015rya} to obtain a very explicit realisation of the KNP alignment mechanism in which the effective axion decay constant can be explicitly deduced from the various fluxes. We will use this explicit construction to attempt to enhance the effective decay constant by inducing alignment by scanning over various flux values. We shall find that there is a lower limit on the alignment angle as well as a sub-Planckian upper limit on the value of effective decay constant for all the flux values we consider.

In \textsection \ref{sec:typeIIA_reminder}, we remind the reader the basic equations relevant for understanding the dynamics of axions in IIA theory. 
Then, in \textsection \ref{sec:enhancement}, we attempt to enhance the effective decay constant, in particular, in \textsection \ref{sec:my_formalism}, we present a method of doing so and study various consequences of this method in the later subsections.
In \textsection \ref{sec:failure}, we find that this method does not lead to the desired results. Thus, in \textsection \ref{sec:careful}, we attempt to obtain an enhanced effective decay constant by an explicit realisation of KNP alignment mechanism in type IIA theory (the details of this construction are given in \textsection \ref{sec:KNP}. We find that this approach also fails to provide a super-Planckian effective decay constant. Finally, in \textsection \ref{sec:discussion}, we conclude with a discussion of various related issues.

\section{A quick reminder of key concepts}

\subsection{Type IIA flux vacua} \label{sec:typeIIA_reminder}

In this work, we shall follow the notations and conventions used in ref \cite{Palti:2015xra}.
It is well known that compactifying type IIA string theory on orientifolds of Calabi-Yau threefolds ($CY_3$) gives rise to ${\cal N}=1$ supergravity theory in 1+3 spacetime dimensions \cite{Grimm:2004ua}. 
We note that in the following, we focus attention to four dimensional effective field theory obtained from massive type IIA supergravity theory in ten dimensions \cite{Romans:1985tz}.

\subsubsection{${\cal N} = 1$ supergravity} \label{sec:sugra}

The dynamics of the scalar sector of ${\cal N}=1$ supergravity theory in 1+3 spacetime dimensions is determined by a Kahler potential and a superpotential (along with other quantities which won't play any role in what follows). 
Recall that if the complex scalar fields in the theory are denoted as $\phi_i$, then, the Kahler potential $K(\phi_i, {\bar \phi}_{\bar j})$ is a real function of these fields and has mass dimension $+2$. Similarly, the superpotential $W(\phi_i)$ is a holomorphic function of the fields and it has a mass dimension $+3$. The Lagrangian for the scalar sector (for those scalars which are not gauged i.e. in the absence of a D-term potential) is given by the expression
\begin{equation} \label{eq:kin}
{\cal L} = K^{i {\bar j}} \partial_\mu \phi_i \partial^\mu {\phi^\dagger}_{\bar j} - V_F \; 
\end{equation}
where, the F-term scalar potential is given by
\begin{equation} \label{eq:pot}
V_F = e^{\frac{K}{M_p^2}} \left[ K^{i {\bar j}} D_i W D_{\bar j} {\bar W} - \frac{3 |W|^2}{M_p^2} \right] \; ,
\end{equation}
note that here, the Kahler covariant derivative is given by $D_i W = \partial_i W + \frac{W \partial_i K}{M_p^2}$ and
$K_{i {\bar j}} = \frac{\partial^2 K}{\partial \phi_i \partial {\bar \phi}_{\bar j}}$, while $K^{i {\bar j}}$ is the inverse of $K_{i {\bar j}}$.

\subsubsection{${\cal N} = 1$ supergravity from IIA: fundamentals}

The ${\cal N} = 1$ supersymmetry of the four-dimensional effective theory would ensure that all scalar fields are complex, and therefore pseudoscalar axions exist along with scalar moduli. For type IIA supergravity on orientifolds of a $CY_3$, we have three sets of complex scalars:
$T_i = b_i + i t_i$ (called  complexified Kahler moduli, here, $i$ runs from 1 to $h^{1,1}_-$), $U_\lambda = u_\lambda + i \nu_\lambda$, where $\lambda = 1,2,\dots,h^{2,1}$ (here, $u_\lambda$ are the complex structure moduli while $\nu_\lambda$ are axions) and, finally, $S = s + i \sigma$ (here, $s$ is the dilaton and $\sigma$ is one of the axions). 
Recall that the vacuum expectation values of the moduli fields determine the shape or size of certain topological cycles in the extra dimensions. 

The Kahler potential for the resulting theory is given by a sum of three contributions
\begin{equation} \label{eq:kahler_pot}
K = - \ln 8 {\cal V} - \ln (S+{\bar S}) - 2 \ln {\cal V}'  \; ,
\end{equation}
here, $\cal V$ depends on the Kahler moduli $t_i$ alone (see \cite{Palti:2015xra} for its exact expression) while ${\cal V}'$ depends on complex structure moduli $u_\lambda$ alone. 
Note that the Kahler potential does not depend on the axions i.e. $b_i$, $\sigma$ and $\nu_\lambda$ due to a perturbative shift symmetry.
In this work, we focus our attention on two simple and quiet similar cases:

(a) $CY_3$ which is mirror of the quintic, for which $h^{2,1}=1$, ${\cal V}' = u_1^{3/2}$ (we will call this the two axion case), 
and, 

(b) $CY_3$ which is mirror to $P_{[1,1,6,9]}$ manifold for which $h^{2,1} = 2$ while ${\cal V}'$ is given by \cite{Denef:2004dm}
\begin{equation} \label{eq:cmplx_vol}
 {\cal V}' = u_1^{3/2} - u_2^{3/2} \; ,
\end{equation}
which we will call the three-axion case. Note that one could obtain $ {\cal V}'$ of two-axion case from the ${\cal V}'$ of three-axion case by setting $u_2 = 0$.

The moduli and axions correspond to flat directions of the scalar potential (evaluated at the leading order) and they are stabilized by fluxes or instantons, to which we now turn.

\subsubsection{Fluxes, superpotential and perturbative moduli stabilization}

In ten dimensions, Type IIA supergravity has the following $p-$form gauge potentials: $B_2$ (and its magnetic dual $B_6$) in NS sector as well as $A_1$, $C_3$ (and their duals $A_7$, $C_5$) in RR sector \cite{Grimm:2004ua}. Thus, the possible field strengths are $H_3 = dB_2$ as well as $F_2 = dA_1$ and $F_4 = d C_3 + \cdots$. In addition, we have two more kinds of fields strength available: (a) since we are working with massive type IIA supergravity, the mass parameter of the theory acts as $0-$form flux $F_0$ \cite{Romans:1985tz}, and (b) there is a $6-$form flux $F_6$, which is basically the volume form on the compact manifold. If one compactifies on a manifold with a non-trivial $(p+1)-$cycle $\Sigma_{p+1}$, one could consider configurations in which the flux of any of the field strengths $F_{p+1}$ could be non-zero on one of these cycles in the extra dimensions. When such fluxes are turned on, a superpotential is induced in the four dimensional effective theory (\cite{Grimm:2004ua,DeWolfe:2005uu,Acharya:2006ne,Palti:2008mg}, also, see \cite{Palti:2015xra} for the details relevant in our context) and the compactification manifold is no longer a Calabi-Yau manifold, but for small enough fluxes, the backreaction of fluxes on the compactification manifold can be ignored. 

One could then find the scalar potential using Eq (\ref{eq:pot}) and look for supersymmetric critical points, by using the condition $D_a W = 0$ \cite{Wess:1992cp}. Notice that since we are looking for supersymmetric critical points, they can not be de-Sitter. By following this procedure, one finds that all the geometric moduli (Kahler moduli and complex structure moduli) get fixed while only one linear combination of RR axions ($\sigma$, $\nu_\lambda$) gets fixed, thus, unless $h^{2,1} = 0$, this leaves some axions unfixed.

For our purpose, the flux values themselves can be thought of as ``free parameters" which we can adjust to obtain different solutions.
When we focus attention on the $(S, U_1, U_2)$ sector of the scalar field space, the corresponding ``free parameters" are the fluxes denoted as $q^1$, $q^2$, $f_0$ and $h_0$ (the dilaton flux parameter) as well as the volume ${\cal V}$ (see \cite{Palti:2015xra} for details). 
In the three-axion case, moduli stabilization at the classical level fixes the vevs of the fields $s$, $u_1$ and $u_2$ to the values determined by the fluxes as follows \cite{Palti:2015xra}
\begin{eqnarray}
s &=& \left( \frac{2 f_0}{5 h_0} \right) {\cal V} \; , \label{eq:s} \\ 
u_1 &=& \left( \frac{(q^1)^2}{(q^1)^3 + (q^2)^3} \right) 3 h_0 s \; , \label{eq:u1} \\
u_2 &=& \left( \frac{(q^2)^2}{(q^1)^3 + (q^2)^3} \right) 3 h_0 s \; . \label{eq:u2} 
\end{eqnarray}
On the other hand, only the following linear combination of the axions is fixed by this procedure.
\begin{equation} \label{eq:condition}
h_0 \sigma + q^1 \nu_1 + q^2 \nu_2 = {\rm constant} \; .
\end{equation}
Here, the two-axion case can be arrived at by setting $\nu_2$ and $u_2$ to zero.

\subsubsection{Non-perturbative effects}

In the saddle point or semiclassical approximation, the amplitude of a process (e.g. tunneling) goes as ${\cal A} \sim e^{-S_E}$, where $S_E$ is the Euclidean action of the process.
There are non-perturbative contributions to the vacuum in string theory \cite{Dorey:2002ik,Blumenhagen:2009qh,Ibanez:2012zz}, whose strength can be found by the following consideration: if the compactification manifold has a topologically non-trivial spacelike p-cycle, $\Sigma^i_p$, this completely spacelike cycle could be wrapped by an Euclidean D(p-1)-brane (called an instanton, since the worldvolume of such a brane is localized in the time-direction of target space just like a gauge instanton is localized in spacetime rather than space).
The contribution of the Euclidean D(p-1)-brane wrapping a p-cycle $\Sigma^i_p$ to the path integral (and hence the strength of the instanton) is then given by (see Eq (13.15) of \cite{Ibanez:2012zz}):
\begin{eqnarray} 
{\cal A} \propto e^{-S_{E_{p-1}}} 
&=& e^{- \frac{2 \pi}{{\ell}_s^p} \left( \frac{1}{g_s} {\rm Vol}(\Sigma^i_p) + i \int_{\Sigma^i_p} C_p \right)} \nonumber\\
&=& e^{- \frac{2 \pi}{g_s} \left(\frac{{\rm Vol}(\Sigma^i_p)}{{\ell}_s^p} \right)} e^{-i a_i} \; ,
\end{eqnarray}
here, ${\ell}_s$ is the string length, $g_s$ is the string coupling, ${\rm Vol}(\Sigma_p^i)$ is the volume of the p-cycle charcterised by the index $i$, $C_p$ is the RR p-form gauge potential, $a_i$ is the $i^{\rm th}$ axion and the factor $i$ comes from the Euclidean signature of the brane. The last term generates a cosine potential for the $i^{\rm th}$ RR axion corresponding to $C_p$. In the case of our interest, the superpotential generated by fluxes could receive non-perturbative corrections from Euclidean D2-brane instantons, which is of the form \cite{Palti:2015xra}
\begin{equation}
W = W_{\rm perturbative} + \sum_I A_I e^{- a_0^I S - a_\lambda^I U_\lambda} \; ,
\end{equation}
Given this generic form of the superpotential and the Kahler potential given in Eq (\ref{eq:kahler_pot}), one can easily find the scalar potential by using Eq (\ref{eq:pot}). For the correct choices of $a^I_0$ and $a^I_{\lambda}$, the potential would be of the form shown in Eq (\ref{eq:Lag_low_2}), Eq (\ref{eq:Lag_low_3}) or Eq (\ref{eq:Lag_low_1}) below. 
Note that the quantities $a^I_0$ and $a^I_{\lambda}$ could in principle be deduced from string theory (they would be the intersection numbers between the 3-cycle wrapped by the instanton and the 3-cycles associated to the $U_{\lambda}$  \cite{Ibanez:2012zz}). 
These choices are thus determined by the choice of the compactification manifold, and since there there is a lot of freedom to choose the compactification manifold, there is a wide range of possibilities for the values of $a^I_0$ and $a^I_{\lambda}$ too. Let us note that, in this simple type IIA setting, the values of the moduli $s$ and $u_\lambda$ can be explicitly determined in terms of fluxes using Eq (\ref{eq:s}), Eq (\ref{eq:u1}) and Eq (\ref{eq:u2}). 

\subsection{Kim-Nilles-Peloso mechanism}

Before proceeding, we'd like to quickly remind the reader of the celebrated Kim-Nilles-Peloso mechanism
 \cite{Kim:2004rp}, which is often also referred to as lattice alignment mechanism and which we will call KNP mechanism from here onwards (see also \cite{Choi:2014rja}). 
 
Consider a two dimensional scalar field space $(\phi_1, \phi_2)$, if the potential $V_0$ is of the form
\begin{equation}
V_0(\phi_1, \phi_2) = \Lambda^4 \left[ 1 - \cos \left( \frac{\phi_1}{f_1} \right) \right] \; ,
\end{equation}
then, obviously, the $\phi_1$ direction in field space is periodic (with period $2 \pi f_1$) while the direction orthogonal to it, the $\phi_2$ direction, is a flat direction (this is equivalent to saying that the corresponding decay constant is infinity). 

Now consider a situation in which the potential is given by 
\begin{equation}\label{eq:pot_tmp_1}
 V_1 = A_1 \cos \left[ \frac{\phi_1}{f_1} + \frac{\phi_2}{g_1} \right] \; .
\end{equation}
If one defines a new field $\psi_1$ by the relation
\begin{equation} \label{eq:dir_1}
 \psi_1 =  \frac{\alpha}{\sqrt{\alpha^2 + \beta^2}} \phi_1 + \frac{\beta}{\sqrt{\alpha^2 + \beta^2}} \phi_2 \; ,
\end{equation}
with $\alpha = f_1^{-1}$ and $\beta = g_1^{-1}$, 
then one can interpret the $\psi_1$ direction in field space as the direction which makes an angle $\theta_1$ w.r.t. the $\phi_1$ direction such that $\cos \theta_1 = \frac{\alpha}{\sqrt{\alpha^2 + \beta^2}}$ etc. Thus, the potential in Eq (\ref{eq:pot_tmp_1}) becomes
\begin{equation}
 V_1 = A_1 \cos \left[ ( \sqrt{\alpha^2 + \beta^2}) \psi_1  \right] \; , 
\end{equation}
so that the direction orthogonal to the $\psi_1$ direction must be a flat direction.
Similarly, if the potential is 
\begin{equation}\label{eq:pot_tmp_2}
 V_2 = A_2 \cos \left[ \frac{\phi_1}{f_2} + \frac{\phi_2}{g_2} \right] \; , 
\end{equation}
then there is a direction $\psi_1'$, which makes an angle of $\theta_1'$ with $\phi_1$ axis and for which 
\begin{equation} \label{eq:dir_1'}
 \psi_1' =  \frac{\gamma}{\sqrt{\gamma^2 + \delta^2}} \phi_1 + \frac{\gamma}{\sqrt{\gamma^2 + \delta^2}} \phi_2 \; ,
\end{equation}
with $\gamma = f_2^{-1}$ and $\delta = g_2^{-1}$ and $\cos \theta_1' = \frac{\gamma}{\sqrt{\gamma^2 + \delta^2}}$.
The potential in Eq (\ref{eq:pot_tmp_1}) becomes
\begin{equation}
 V_2 = A_2 \cos \left[ ( \sqrt{\gamma^2 + \delta^2}) \psi_1'  \right] \; , 
\end{equation}
and the direction orthogonal to $\psi_1'$ direction is a flat direction.

Finally, consider a system whose scalar potential is of the form $V = V_1 + V_2$, i.e.
\begin{eqnarray} 
\label{eq:pot_KNP}
V (\phi_1, \phi_2) = \Lambda_1^4 \left[ 1 - \cos \left( \frac{\phi_1}{f_1} + \frac{\phi_2}{g_1} \right)  \right] + \nonumber\\
\Lambda_2^4 \left[ 1 - \cos \left( \frac{\phi_1}{f_2} + \frac{\phi_2}{g_2} \right)  \right] \; .
\end{eqnarray}
In this case, we have two directions in field space, the $\psi_1$ direction, Eq (\ref{eq:dir_1}), and $\psi_1'$ direction, Eq (\ref{eq:dir_1'}), and the angle between these directions can be found by evaluating the cross product of the unit vectors in those directions. We find that
\begin{equation} \label{eq:delta_theta}
\sin \Delta \theta = \frac{(g_1 f_2 - g_2 f_1)}{\sqrt{(f_1^2 + g_1^2)(f_2^2 + g_2^2)}} \; .
\end{equation}
It is clear that, when $\Delta \theta = 0$ i.e. when $\theta_1 = \theta_1'$, i.e. the directions $\psi_1$ and $\psi_1'$ are aligned, then, the direction orthogonal to $\psi_1$ (and $\psi_1'$) will be a flat direction in field space. Now, $\theta_1 = \theta_2$ implies that $\tan \theta_1 = \tan \theta_2$, i.e. 
\begin{equation}
\frac{f_1}{g_1} = \frac{f_2}{g_2} \; .
\end{equation}
We may wish to avoid having an exactly flat direction in field space (because that would correspond to having a massless scalar field in the spectrum), so, we may be interested in the situations in which there is a sufficiently light scalar whose mass is small in Planck units but big enough to avoid any detection.
The most important insight is that under certain conditions, even when the above condition does not get satisfied exactly, there exists a direction in field space along which $f$ is large.

Consider a field redefinition such that the first field direction, $\psi_1$ is still given by Eq (\ref{eq:dir_1}) and the second field direction, $\psi_2$ is orthogonal to $\psi_1$. One can rewrite the potential Eq (\ref{eq:pot_KNP}) in terms of these new fields and obtain
\begin{eqnarray}
V(\psi_1,\psi_2) = \Lambda_1^4 \left[ 1 - \cos \left( \frac{\psi_1}{f_1'} \right)  \right] + \nonumber\\
\Lambda_2^4 \left[ 1 - \cos \left( \frac{\psi_1}{f_2'} + \frac{\psi_2}{f_{\rm eff}} \right)  \right] \; ,
\end{eqnarray}
where,
\begin{eqnarray}
 f_1' &=& \frac{f_1 g_1}{\sqrt{f_1^2 + g_1^2}} \; , \\
 f_2' &=& \frac{f_2 g_2 \sqrt{f_1^2 + g_1^2}}{f_1 f_2 + g_1 g_2} \; , \\
  f_{\rm eff} &=& \frac{f_2 g_2 \sqrt{f_1^2 + g_1^2}}{f_1 g_2 - g_1 f_2} \; . \label{eq:feff}
\end{eqnarray}
If one could freely choose $f_1, f_2, g_1, g_2$ such that $f_1 g_2 - g_1 f_2$ becomes too small, then, $m^2_{\psi_2} \ll m^2_{\psi_1}$, and one can set $\langle \psi_1 \rangle = 0$, one would then get 
\begin{eqnarray}
V(\psi_2) \approx \Lambda_2^4 \left[ 1 - \cos \left( \frac{\psi_2}{f_{\rm eff}} \right)  \right] \; ,
\end{eqnarray}
with $f_{\rm eff}$ chosen to be as large as desired.
Thus, if one sits at the origin in field space, there exists a direction (the heavier, $\psi_1$ direction) in which the potential rises very steeply and there exists a direction orthogonal to this (the lighter, $\psi_2$ direction) in which the potential rises very slowly. 

This KNP alignment mechanism has been studied in the context of string theory also (see e.g. \cite{Long:2014dta,Gao:2014uha} and \cite{Hebecker:2018fln}).
In ref \cite{Palti:2015xra}, the author, inspired by \cite{Hebecker:2015rya}, studied a way to realise a rough version of KNP mechanism for type IIA string vacua. 
Recall that in this case, by simply turning the fluxes on, one can generate a potential which would fix all Kahler moduli, complex structure moduli and a linear combination of axions. 
In \cite{Palti:2015xra}, the most explicitly studied case is the one with two-dimensional axion field space: there is one heavy direction and in the direction orthogonal to it, which is flat at the perturbative level, the potential is generated by non-perturbative effects (and hence, is a cosine). It was then argued that, unlike Eq (\ref{eq:feff}), the decay constant in this stringy set up can not be enhanced to arbitrarily large values.
It is this feature of the model studied in \cite{Palti:2015xra} that we wish to study in greater detail.
{\footnote{
Before proceeding, we note that in ref \cite{Palti:2015xra}, the author focusses on a case in which $\sigma$, the superpartner of dilaton and $u_1$, the superpartner of largest modulus are the only ones being mixed. Even in the case in which there are three axions, the author adjusts fluxes to ensure that the axion which is the superpartner of the smaller modulus becomes sufficiently heavy that it can be ignored from the low energy dynamics.}
}
We will do so in two different ways, in particular, in \textsection \ref{sec:KNP}, we will arrive at a very explicit realisation of the KNP alignment mechanism.

\section{First attempt to obtain large effective decay constant} \label{sec:enhancement}

In this section, we shall attempt to enhance axion decay constant in the set up of IIA theory presented in the last section.
This problem was studied in ref \cite{Palti:2015xra} which we closely follow. As we shall see, there are important new lessons to be learnt even in the simple case of a $CY_3$ which is mirror to $P_{[1,1,6,9]}$ manifold \cite{Denef:2004dm}, for which $h^{2,1} = 2$ and there are three scalar fields. 

To begin with, however, we remind ourselves of how to deal with a slightly different situation, the two-axion case mentioned in the last section i.e. mirror of the quintic for which $h^{2,1} = 1$ and ${\cal V}' = u_1^{3/2}$.
This case has already been studied in \cite{Palti:2015xra}, but as we shall see, there are important observations to be made in order to study the more interesting case of mirror to $P_{[1,1,6,9]}$ with $h^{2,1} = 2$ i.e. the three axion case.
In \cite{Palti:2015xra}, the author only briefly mentions the three axion case (in particular, in \cite{Palti:2015xra} only the two-axion limit of the three axion case is mentioned). 
In the upcoming subsection, we revisit the two axion case while in the sub section after that, we analyse the three axion case.

\subsection{The two-axion (i.e. $h^{2,1} = 1$) case}

Recall that in type IIA theory, moduli stabilization at leading order in $\alpha'$ and $g_s$ ensures that only a single linear combination of axions is fixed \cite{Grimm:2004ua,DeWolfe:2005uu}. 
When we have only two axions, this defines a unique straight line in field space which is a flat direction at perturbative level. If distance along this direction is thought of as a field, this field is perturbatively massless. As we saw, non-perturbative effects such as Euclidean brane instantons will then lift this flat direction and generate a potential for the perturbatively massless field. 

In this case, the fluxes we could vary (to understand axion dynamics) are $q^1, f_0, h_0$ and we could think of the volume of the compactification manifold $\cal V$ as another ``free parameter." The values of moduli $s$ and $u_1$ can be found in terms of these variables.

\subsubsection{Basics}

Here, ${\cal V}'$ can be obtained from Eq (\ref{eq:cmplx_vol}) by setting $u_2=0$ and so, using the equations presented in \textsection  \ref{sec:sugra}, we can show that
\begin{equation} \label{eq:k_metric_2}
K_{S {\bar S}} = \frac{1}{4 s^2} \; ,~ K_{U_1 {\bar U}_1} = \frac{3}{4 u_1^2} \; ,
\end{equation}
This is the metric in $S-U_1$ field space and this will determine the kinetic terms of $s, \sigma, u_1$ and $\nu_1$.
Recall that fluxes generate a potential and hence fix the values of the moduli $s$ and $u_1$ while for the axions $\sigma$ and $\nu_1$, a linear combination viz. $h_0 \sigma + q^1 \nu_1$, gets fixed. Thus, in the $\sigma-\nu_1$ field space, there exists a direction which is flat at perturbative level. At low energies, we can think of $s$ and $u_1$ as essentially fixed quantities.

The Euclidean D2-brane instantons could generate a potential for the axions. The actual form of the potential would depend on the details such as which cycles the brane wraps, this determines the instanton. There must be solutions in which the instantons happen to be such that the Lagrangian determining the dynamics of the remaining low energy fields (i.e. $\sigma, \nu_1$) is given by (see e.g. \cite{Palti:2015xra})
\begin{eqnarray} \label{eq:Lag_low_2}
{\cal L} = &-&\frac{1}{2} f_\sigma^2 (\partial \sigma)^2 - \frac{1}{2} f_{\nu_1}^2 (\partial \nu_1)^2 
 - \Bigl[V_0 + A' e^{-s} (1-\cos \sigma) \nonumber \\
&+& B' e^{-u_1} (1- \cos \nu_1) 
\Bigr],
\end{eqnarray}
where, the potential for the axions is generated by non-perturbative effects. 
On comparing Eq (\ref{eq:kin}), Eq (\ref{eq:k_metric_2}) and Eq (\ref{eq:Lag_low_2}), we find that $f_\sigma$ is dependent on $s$ while $f_{\nu_1}$ is dependent on $u_1$ i.e.
\begin{equation} \label{eq:def_fs}
 f_\sigma = \frac{1}{\sqrt{2}s} \; ,~ f_{\nu _1} = \frac{\sqrt 3}{\sqrt{2} u_1} \; .
\end{equation}
Needless to say, the canonically normalised axions are $f_\sigma \sigma$ and $f_{\nu_1} \nu_1$. 
Before proceeding, we note the following:

\begin{enumerate}
 \item The advantage of restricting our attention to instantons which lead to the potential shown in Eq (\ref{eq:Lag_low_2}) is that  the periods as well as the amplitudes of the cosines are known in terms of moduli. 
Finally, note that, in this model, we also explicitly know the values of moduli themselves in terms of fluxes.
 \item When one ignores the non-perturbative effects, at leading order in $\alpha'$ and $g_s$, the linear combination $h_0 \sigma + q^1 \nu_1$ is fixed to some value. By redefining the fields, one could ensure that
\begin{equation} \label{eq:line}
h_0 \sigma + q^1 \nu_1 = 0 \; .
\end{equation}
Notice that if we keep the flux $h_0$ fixed and increase the flux $q^1$, then, the straight line representing perturbative flat direction tends to align with the axion $\nu_1$.
 \item The moduli $s$ and $u_1$ have geometrical interpretation, they are related to volumes of certain topological cycles (in units of string length). Thus, if we wish to stay in the trustworthy regime of effective field theory, we must have $s > 1$ and $u_1 > 1$.
From the expressions for the decay constants $f_\sigma$ and $f_{\nu _1}$ in Eq (\ref{eq:def_fs}), it is then clear that these decay constants have to be smaller than one in Planck units.
\end{enumerate}

In the $(f_\sigma \sigma, f_{\nu_1} \nu_1)$ plane of the canonically normalised fields, this describes a straight line passing through the origin i.e.
\begin{equation} \label{eq:line_2}
\left( \frac{h_0}{f_\sigma} \right) f_\sigma \sigma + \left( \frac{q^1}{f_{\nu_1}} \right) {f_{\nu_1}} \nu_1 = 0 \; .
\end{equation}
The slope of this line is given by $-(h_0 f_{\nu_1})/(f_\sigma q^1)$ and the two direction cosines of the line are
\begin{eqnarray}
{\ell}_\sigma &=& \frac{q_1 f_\sigma}{N} \; , \\
{\ell}_{\nu_1} &=& - \left(\frac{h_0 f_{\nu_1}}{N} \right) \; ,
\end{eqnarray}
where, $N = \sqrt{f_\sigma^2 (q^1)^2 + f_{\nu_1}^2 (h_0)^2}$. 
At this stage, it is worth recalling that in $N-$dimensional Euclidean space with Cartesian coordinates $(x_1,x_2,\dots,x_N)$,
the distance $r$ along any straight line passing through the origin (and with direction cosines $({\ell}_1,\dots,{\ell}_N)$) is 
$r = {\ell}_1 x_1 + \dots + {\ell}_N x_N$. If we now call $\psi$ to be the distance along the direction described by line Eq (\ref{eq:line_2}), one finds that 
\begin{equation} \label{eq:psi}
 \sigma = \left( \frac{q^1}{N} \right) \psi \; , ~~  \nu_1 = \left( \frac{-h_0}{N} \right) \psi \; .
\end{equation}
If we go along the straight line direction Eq (\ref{eq:line_2}), non-perturbative effects shall generate a potential which can be found by substituting for $\sigma$ and $\nu_1$ from Eq (\ref{eq:psi}) into Eq (\ref{eq:Lag_low_2}), one thus obtains, 
\begin{eqnarray} \label{eq:Lag_low_2_psi}
{\cal L} = -\frac{1}{2} (\partial \psi)^2 - \Biggl[ V_0' + A' e^{-s} \left(1-\cos \frac{\psi}{ f_\psi^{s}} \right) \nonumber \\
+ B' e^{-u_1} \left(1- \cos \frac{\psi}{ f_\psi^{u_1}} \right) \Biggr] \; ,
\end{eqnarray}
where, $f_\psi^{s} = N/q^1$ and $f_\psi^{u_1} = N/h_0$. Since in the low energy theory we can think of $s$ and $u_1$ as fixed quantities, the potential experienced if one moves along the straight line direction Eq (\ref{eq:line_2}) is a function of only one field, the distance $\psi$ along this direction. Then, Eq (\ref{eq:Lag_low_2_psi}) suggests that the potential of $\psi$ is a sum of two cosines with different amplitudes and periods. 

\subsubsection{Flux independence of the slope of fixed direction} \label{sec:flux-ind}

The coefficients of $\sigma$ and $\nu_1$ in Eq (\ref{eq:line}) are clearly flux dependent, hence, by changing the fluxes, we could change the slope of the line in $\sigma-\nu_1$ plane. Now consider the line in $(f_\sigma \sigma, f_{\nu_1} \nu_1)$ plane of the canonically normalised fields, the slope of the straight line in Eq (\ref{eq:line_2}) is $-(h_0 f_{\nu_1})/(f_\sigma q^1)$. Using Eq (\ref{eq:def_fs}), eq (\ref{eq:s}) and eq (\ref{eq:u1}) (with $q^2$ set to 0), we find that this slope is equal to $-1/{\sqrt 3}$. I.e the fixed direction in the space of canonically normalised fields makes an angle $-\pi/6$ w.r.t. the positive $f_\sigma \sigma$ axis.
Thus, by changing the fluxes, we can not change the orientation of the straight line in the plane of canonically normalised fields.

\subsubsection{Obstruction to flat potential} \label{sec:obstruction_2}

Now, in this context one could think about the potential along the $\psi$ direction and its possible flatness. One of the things we mean when we say that the potential along the straight line direction Eq (\ref{eq:line_2}) is pretty flat is that it is a cosine with very large period. Suppose that one of decay constants among $f_\psi^{s}$ and $f_\psi^{u_1}$, say the latter, is very large and that $s$ is large as compared to $u_1$, then, the amplitude of the first cosine in Eq (\ref{eq:Lag_low_2_psi}) is exponentially suppressed as compared to the second cosine while the period of this second cosine is also large, thus, we could get a direction in which the potential is quite flat.
Note that,
\begin{eqnarray}
 f_\psi^{s} = \frac{N}{q^1} = \frac{\sqrt{f_\sigma^2 (q^1)^2 + f_{\nu_1}^2 (h_0)^2}}{q^1}, 
  ~s = \frac{2 f_0 {\cal V}}{5 h_0} \; ,\\
 f_\psi^{u_1} = \frac{N}{h_0} = \frac{\sqrt{f_\sigma^2 (q^1)^2 + f_{\nu_1}^2 (h_0)^2}}{h_0}, 
  ~u_1 = \frac{3 h_0 s}{q^1} \; .
\end{eqnarray}
Thus, it appears that if one keeps $h_0$, $f_0$ and $\cal V$ fixed and increases $q^1$, then for large enough $q^1$, $s$ stays put while $u_1$ decreases and $f_\psi^{s}$ stays constant while $f_\psi^{u_1}$ increases. Thus, one might conclude that the first cosine in Eq (\ref{eq:Lag_low_2_psi}) shall become suppressed over the second one while the period of the second one could be made large, thus, flattening the potential. Furthermore, using eq (\ref{eq:def_fs}),
eq (\ref{eq:s}) and eq (\ref{eq:u1}) (with $q^2$ set to 0), we conclude that
\begin{equation}
 f_\psi^{u_1} = \frac{q^1}{{\sqrt 3} h_0 s} \; ,
\end{equation}
so that increasing $q^1$ with fixed $s$ (by holding $h_0$, $f_0$ and $\cal V$ fixed) will cause 
$f_\psi^{u_1}$ to increase as much as we like without any consequences. 

This happens to be not true, since it turns out that
\begin{eqnarray} \label{eq:2-axion-eff-decay}
 f_\psi^{s} = \frac{1}{{\sqrt 3} s} = \frac{\sqrt{2} f_\sigma}{\sqrt 3} \; , \\
 f_\psi^{u_1}  = \frac{\sqrt 3}{u_1} = \sqrt{2} f_{\nu_1} \; . \label{eq:2-axion-eff-decay_2}
\end{eqnarray}
Since $ f_\psi^{s}$ and $ f_\psi^{u_1}$ are simply proportional to the fundamental axion decay constants $f_\sigma$ and $f_{\nu_1}$, and since these fundamental decay constants can not be super-Planckian, we conclude that $ f_\psi^{s}$ and $ f_\psi^{u_1}$ shall also remain sub-Planckian. 
even though we could increase $ f_\psi^{s}$ and $f_\psi^{u_1}$, we can not make them so large that $u_1$ and $s$ become too small. Since $s$ and $u_1$ are geometric moduli which determine the sizes and shapes of compactification manifold, they can not be made too small without leaving the regime of validity of low energy effective field theory.

For our purpose, we note that the factors relating $ f_\psi^{s}$ to $f_\sigma$ and $ f_\psi^{u_1}$ to $f_{\nu_1}$ are ${\cal O}(1)$ numbers. We shall see that in two-axion limit of three-axion case, there is additional freedom which can cause these factors to be very large numbers.

\subsection{The three axion case} \label{sec:3axion}

Going beyond the work of ref \cite{Palti:2015xra}, in this subsection, we shall analyse the case with three axions and recover the two-axion case as a limiting case. It would appear that the three axion case offers new features and there is scope for enhancement of decay constant. But as we shall see at the end of this subsection, this is not so. 

For the case with three axions, the Lagrangian of the low energy effective theory contains three scalar fields $\sigma, \nu_1$ and $\nu_2$. In the absence of non-perturbative corrections, the condition given by Eq (\ref{eq:condition}) is satisfied, this defines a plane $P$ in $\sigma-\nu_1-\nu_2$ space, which will call the perturbatively flat plane. In the absence of non-perturbative effects, the scalar potential along this plane is constant. But, for small enough $g_s$ and large enough compactification volume, the potential due to the non-perturbative effects would be quite small compared to the one generated by fluxes, thus, the energy required to leave the plane will be far larger than the potential generated by non-perturbative effects. Thus, at low energies, one would be ``stuck" in this plane.

In plane $P$, there are multiple straight line directions one could go to and each of these can be thought of as the field of interest. The question we ask is, can we get the potential experienced along any such direction as a cosine with a large period i.e. is sufficiently flat? 
As we shall see, it appears that this can be done. Notice that we have no guarantee that the field will actually go along such a straight line trajectory (though its dynamics is determined by, among other things, its potential). In any case, we'd argue at the end of this subsection that this approach shall not work.

Given the Kahler potential, the metric in the scalar field space can be found from 
\begin{equation}
K_{i \bar{j}} = \left(\frac{\partial^2 K}{\partial U_i \partial {\bar U}_{\bar j}}\right) \; ,
\end{equation}

which in the $(U_1,U_2)$ subspace of the scalar field space turns out to be
\begin{eqnarray} \label{eq:kahler-metric}
K_{i \bar{j}} 
 =  \alpha \left(\begin{array}{cc} 6u_1 + \frac{3 u_2^{3/2}}{u_1^{1/2}} & -9 \sqrt{u_1 u_2}
\\ -9 \sqrt{u_1 u_2} & 6u_2 + \frac{3 u_1^{3/2}}{u_2^{1/2}} \end{array}\right) ,
\end{eqnarray}
where, $\alpha = \frac{1}{8 \left( u_1^{3/2} - u_2^{3/2} \right)^2}$.
We now restrict our attention to the subspace of $(U_1,U_2)$ which is spanned by $(\nu_1, \nu_2)$. In this two dimensional subspace, notice that the metric still depends on the vev of the moduli $u_1$ and $u_2$. 

Given the Kahler metric $K_{i {\bar j}} (u_1, u_2)$, we could find its eigenvalues (which we call $f_{{\tilde \nu}_1}^2$ and $f_{{\tilde \nu}_2}^2$) and eigenvectors. If one performs a change of basis such that the eigenvectors are used as the basis vectors, then the metric in the new basis is diagonal. One can then make an additional anisotropic scaling transformation to turn the metric into an identity matrix. 
Let the normalised eigenvectors of the metric be denoted by ${\tilde \nu_1}$ and ${\tilde \nu}_2$ and let $P$ be the matrix of change of basis from $(\nu_1, \nu_2)$ to $({\tilde \nu_1}, {\tilde \nu}_2)$, i.e.
\begin{equation} \label{eq:trans}
\nu_i = P_{ij} {\tilde \nu_j} \; ,
\end{equation} 
since the metric is real-symmetric, $P$ must be an orthogonal transformation. 
One must note that all these quantities depend on the moduli $(u_1, u_2)$ which themselves depend on the fluxes.

\subsubsection{Search directions and enhancement} \label{sec:my_formalism}

Perturbative moduli stabilisation ensures that at low energies we stay stuck in a plane, plane $P$, in the $(\sigma,\nu_1,\nu_2)$ space.

\begin{equation}
h_0 \sigma + q^1 \nu_1 + q^2 \nu_2 = 0 \; ,
\end{equation}
Using Eq (\ref{eq:trans}) and after scaling, this implies that
\begin{eqnarray} \label{eq:plane}
\left(\frac{h_0}{f_\sigma} \right) (f_\sigma \sigma) &+& \left(\frac{q^1 P_{11} + q^2 P_{21}}{f_{\tilde \nu_1}} \right) (f_{\tilde \nu_1} {\tilde \nu_1}) \nonumber \\
&+& \left(\frac{q^1 P_{12} + q^2 P_{22}}{f_{\tilde \nu_2}} \right) (f_{\tilde \nu_2} {\tilde \nu_2}) = 0 \; ,
\end{eqnarray}
where, we have simply rewritten the previous equation in terms of normalised eigenvectors of the Kahler metric. This normalisation off-course also canonically normalises the axions we work with i.e. the fields $f_\sigma \sigma$, $f_{\tilde \nu_1} {\tilde \nu_1}$ and $f_{\tilde \nu_2} {\tilde \nu_2}$ are the canonically normalised axions. Needless to say, in the above equation $f_{\sigma}$ is a function of $s$ while $f_{\tilde \nu_i}$ and $P_{ij}$ are functions of $(u_1, u_2)$. 

\begin{figure}
  \includegraphics[width = 0.55\textwidth]{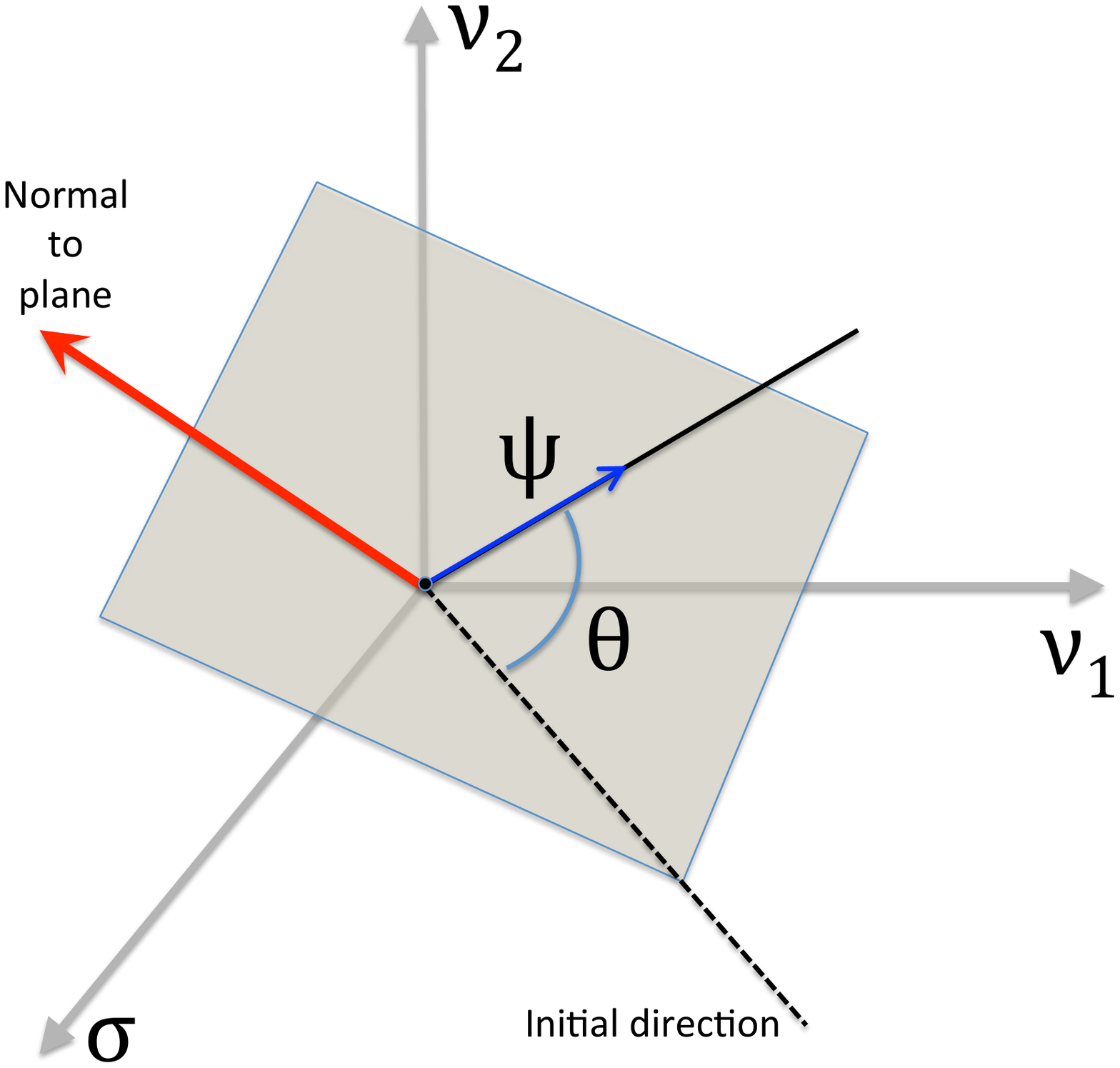}
  \caption
 {This figure provides a picture of the perturbatively flat plane, the normal to the plane, the initial search direction and the search direction for a chosen angle $\theta$ in the $\sigma-\nu_1-\nu_2$ field space.
 }
  \label{diagram}
\end{figure}

Now, Eq (\ref{eq:plane}) describes a plane in the $(f_\sigma \sigma, f_{\tilde \nu_1} {\tilde \nu_1}, f_{\tilde \nu_2} {\tilde \nu}_2)$ space of canonically normalised fields and from its defining equation, one can easily read off the components of the unit vector normal to the plane. Consider the line common between the plane Eq (\ref{eq:plane}) and the plane ${\tilde \nu}_2 = 0$. Obviously, the equation of this line is given by 
\begin{equation} \label{eq:initial_search_dir}
\left(\frac{h_0}{f_\sigma} \right) (f_\sigma \sigma) + \left(\frac{q^1 P_{11} + q^2 P_{21}}{f_{\tilde \nu_1}} \right) (f_{\tilde \nu_1} {\tilde \nu_1}) = 0 \; .
\end{equation}
This is a direction in $(f_\sigma \sigma, f_{\tilde \nu_1} {\tilde \nu_1})$ plane and one could go along this direction and ask whether the potential generated by non-perturbative effects could be sufficiently flat. In order to explore the other search directions, one could begin with a unit vector along the line given by the above equation and make a rotation by an angle $\theta$ about the axis which is normal to the plane, see fig (\ref{diagram}). For any choice of this angle $\theta$, there will be a new search direction. Let the direction cosines of this new search direction be $({\ell}_\sigma,{\ell}_{\tilde \nu_1},{\ell}_{\tilde \nu_2})$, notice that these direction cosines depend on $\theta$ in addition to depending on the fluxes. Since the search direction lies in the plane described by Eq (\ref{eq:plane}), its direction cosines must satisfy the equation of the plane (since the plane passes through the origin)
\begin{eqnarray} 
\left(\frac{h_0}{f_\sigma} \right) {\ell}_\sigma &+& \left(\frac{q^1 P_{11} + q^2 P_{21}}{f_{\tilde \nu_1}} \right) {\ell}_{\tilde \nu_1} \nonumber \\
&+& \left(\frac{q^1 P_{12} + q^2 P_{22}}{f_{\tilde \nu_2}} \right) {\ell}_{\tilde \nu_2} = 0 \; .
\end{eqnarray}
Now, let $\psi$ be the distance along the search direction, then, since $({\ell}_\sigma,{\ell}_{\tilde \nu_1},{\ell}_{\tilde \nu_2})$ are direction cosines (recall, discussion just before eq (\ref{eq:psi})), 
\begin{eqnarray}
\label{eq:lsig} {\ell}_\sigma \psi &=& f_\sigma \sigma \; , \\
 {\ell}_{\tilde \nu_1} \psi &=& f_{\tilde \nu_1} {\tilde \nu_1} \; , \\
 {\ell}_{\tilde \nu_2} \psi &=& f_{\tilde \nu_2} {\tilde \nu_2} \; .
\end{eqnarray}
Now, re-expressing the above relations in terms of the original axions $\nu_1$, $\nu_2$ tells us that
\begin{eqnarray}
(P^{-1})_{11} \nu_1 + (P^{-1})_{12} \nu_2 &=& \left( \frac{{\ell}_{\tilde \nu_1}}{f_{\tilde \nu_1}} \right) \psi \; , \\
(P^{-1})_{21} \nu_1 + (P^{-1})_{22} \nu_2 &=& \left( \frac{{\ell}_{\tilde \nu_2}}{f_{\tilde \nu_2}} \right) \psi \; ,
\end{eqnarray}
the above two equations can be used to solve for $\nu_1$ and $\nu_2$ in terms of $\psi$, thus one gets (using the orthogonality of $P$)
\begin{eqnarray}
\label{eq:sigma-3} \sigma &=&  \left( \frac{{\ell}_\sigma}{f_\sigma} \right) \psi \; , \\
\label{eq:nu-1-3} \nu_1 &=& \left[  \frac{P_{22} {\ell}_{\tilde \nu_1} f_{\tilde \nu_2} - P_{21} {\ell}_{\tilde \nu_2} f_{\tilde \nu_1} }{{\rm det}P~ f_{\tilde \nu_1} f_{\tilde \nu_2}} \right] \psi \; , \\
\label{eq:nu-2-3} \nu_2 &=& \left[  \frac{P_{11} {\ell}_{\tilde \nu_2} f_{\tilde \nu_1} - P_{12} {\ell}_{\tilde \nu_1} f_{\tilde \nu_2} }{{\rm det}P~ f_{\tilde \nu_1} f_{\tilde \nu_2}} \right] \psi \; .
\end{eqnarray}
One could assume that for some Euclidean D-brane instantons, after the diagonalisation of Kahler metric, the low energy effective theory is given by 
\begin{eqnarray} \label{eq:Lag_low_3}
{\cal L} = &-& \frac{1}{2} f_\sigma^2 (\partial \sigma)^2 -\frac{1}{2} f_{{\tilde \nu}_1}^2 (\partial {\tilde \nu}_1)^2 -\frac{1}{2} f_{{\tilde \nu}_2}^2 (\partial {\tilde \nu}_2)^2 \nonumber \\
&-& \Bigl[ V_0 + A' e^{-s} (1-\cos \sigma) + B' e^{-u_1} (1- \cos {\tilde \nu}_1) \nonumber \\
&+& C' e^{-u_2} (1- \cos {\tilde \nu}_2) \Bigr] \; ,
\end{eqnarray}
which, when expressed in terms of the field $\psi$, will give the potential
\begin{eqnarray} \label{eq:Lag_low_3_psi}
&& V = \Biggl[ V_0' + A' e^{-s} \left(1-\cos \frac{\psi}{ f_\psi^{s}} \right) \nonumber \\
&& + B' e^{-u_1} \left(1- \cos \frac{\psi}{ f_\psi^{u_1}} \right) 
+ C' e^{-u_2} \left(1- \cos \frac{\psi}{ f_\psi^{u_2}} \right) \Biggr] \; , \nonumber \\
\end{eqnarray}
Comparing Eq (\ref{eq:Lag_low_3}) with Eq (\ref{eq:Lag_low_3_psi}) and using Eqs (\ref{eq:sigma-3}), (\ref{eq:nu-1-3}), (\ref{eq:nu-2-3}), one can thus read-off the effective decay constants,
\begin{eqnarray}
\label{eq:fs_f} f_\psi^{s} &=& \left( \frac{f_\sigma}{{\ell}_{\sigma}} \right) \; , \\
\label{eq:f1_f} f_\psi^{u_1} &=& \left[  \frac{{\rm det}P~ f_{\tilde \nu_1} f_{\tilde \nu_2}}{P_{22} {\ell}_{\tilde \nu_1} f_{\tilde \nu_2} - P_{21} {\ell}_{\tilde \nu_2} f_{\tilde \nu_1}}  \right] \; , \\
\label{eq:f2_f} f_\psi^{u_2} &=& \left[  \frac{{\rm det}P~ f_{\tilde \nu_1} f_{\tilde \nu_2}}{P_{11} {\ell}_{\tilde \nu_2} f_{\tilde \nu_1} - P_{12} {\ell}_{\tilde \nu_1} f_{\tilde \nu_2} } \right] \; .
\end{eqnarray}
This set of equations tell us that in $(f_\sigma \sigma, f_{\tilde \nu_1} {\tilde \nu_1}, f_{\tilde \nu_2} {\tilde \nu}_2)$ space of canonically normalised fields, if we go along a direction with direction cosines $({\ell}_\sigma,{\ell}_{\tilde \nu_1},{\ell}_{\tilde \nu_2})$ and if the distance travelled is the field $\psi$, the potential experienced is given by Eq (\ref{eq:Lag_low_3_psi}), where, the three effective axion decay constants 
$f_\psi^{s}, f_\psi^{u_1}$ and $f_\psi^{u_2}$
are given by the above equation.
It is worth noting that in the above Eq, the matrix elements of $P$ depend on the fluxes while, as mentioned above, the direction cosines depend on fluxes as well as $\theta$. An important questions worth answering is could there be choices of fluxes and $\theta$ which enhance the effective decay constants? 

\subsubsection{A few useful remarks} \label{sec:remarks}

When our search direction is perpendicular to $\sigma$ axis, we are in the region of field space where $\sigma = 0$ and the scalar potential does not depend on $\sigma$. Perpendicularity to $\sigma$ axis also implies that $\ell_\sigma$ is zero. 
So, in Eq (\ref{eq:lsig}), on LHS, $\ell_\sigma = 0$ and on RHS, $\sigma = 0$ (as we are in the plane perpendicular to $\sigma$ axis. In such a case, Eq (\ref{eq:lsig}) becomes indeterminate and we do not expect to find $f_\psi^{s}$ from Eq (\ref{eq:fs_f}).
Similarly, it is possible that for a fixed choice of fluxes, we happen to be exploring a direction such that the denominator in Eq (\ref{eq:f1_f}) or Eq (\ref{eq:f2_f}) becomes zero. Leaving such special cases where the denominator vanishes exactly, one could still ask whether there can be an enhancement of the effective decay constants.

\begin{figure}
  \includegraphics[width = .475\textwidth]{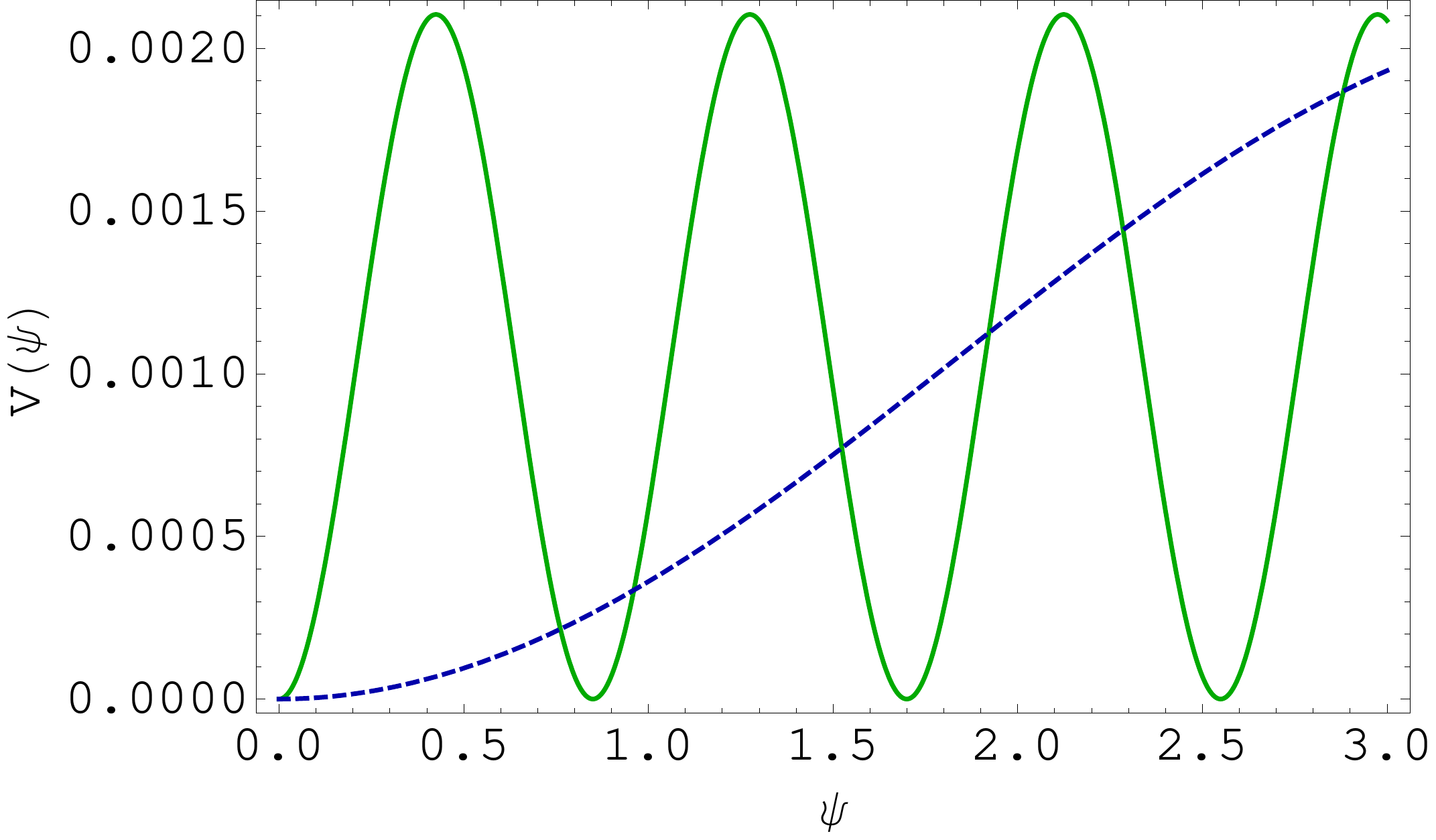}
  \caption
 {For $q^1 = 60$, $q^2 = 40$, $h_0 = 10$, $f_0 = 10$ and ${\cal V} = 100$, one obtains $s \approx 40$, $u_1 \approx 6.86$, $u_2 \approx 15.43$, the resulting axion masses are $m_\sigma \approx 1.65 \times 10^{-7}$,  $m_{\nu_1} \approx 9.83 \times 10^{-1}$ and $m_{\nu_2} \approx 3.8 \times 10^{-3}$. The green curve in this plot is the potential in the direction in $\sigma-\nu_1$ plane while the dashed (blue) curve is the potential in another direction chosen such that there is substantial enhancement in $f_\psi^{u_1}$.
}
  \label{flattened_pot}
\end{figure}

Following the discussion at the beginning of \textsection \ref{sec:obstruction_2}, an important point worth noting is that in Eq (\ref{eq:Lag_low_3_psi}), even if one of the decay constants, say $f_\psi^{u_1}$ is large enough, in order to have a flat potential,  we must also ask whether $s$ and $u_2$ are large enough that the contribution of their potentials (which will be relatively more oscillatory since their decay constants are smaller) in the complete potential would be unimportant. If this can not be ensured, then, even if one of the decay constants, say $f_\psi^{u_1}$ is large, we won't get a flat potential.
Suppose we choose the flux values such that e.g. $u_2$ and $s$ are sufficiently large as compared to $u_1$, then the potential will be mostly dominated by the axion $\nu_1$. For such a fixed choice of fluxes, one could go along any direction in field space (starting from the origin). If the direction cosines of the search direction happen to be such that the denominator in Eq (\ref{eq:f1_f}) becomes small, then, we could have an enhancement of $f_\psi^{u_1}$ as well as get an actual flat potential.
From Eq (\ref{eq:Lag_low_3}), it is easy to see that the mass of each axion would be given by 
\begin{equation}
m^2_i \sim \frac{e^{-u_i}}{f_i^2} \; , 
\end{equation}
and typically, $f_i \sim 1/u_i$, thus, $m^2_i \sim u_i^2 e^{-u_i}$, thus, large vev shall make the axions light (because of the exponential factor). Thus, it is conceivable that the potential can be flattened by this procedure. In fig (\ref{flattened_pot}), we have shown an example of this phenomenon.

Starting from the formalism of three-axion case, one should be able to recover the two-axion case in some limit. This limiting case was briefly mentioned in \cite{Palti:2015xra} but we will find new effects not studied there.
When one chooses $q^1 \gg q^2$, one finds that $u_1 \gg u_2$ and hence $m_{\nu_1} \ll m_{\nu_2}$. 
If the other fluxes are also adjusted to also ensure that $m_\sigma \ll m_{\nu_2}$, then, the axion $\nu_2$ becomes too heavy. We then expect that we should be able to integrate out this heavy axion and recover the two-axion case in a limit.
As we shall see, though this is true, there exist interesting subtleties. To take $q^1 \gg q^2$ limit, we define
\begin{equation}
 \epsilon = \sqrt{\frac{u_2}{u_1}} = \frac{q^2}{q^1} \; ,
\end{equation}
the desired limit is then $\epsilon \rightarrow 0$ limit. Then, in terms of $\epsilon$, the Kahler metric in Eq (\ref{eq:kahler-metric}) takes the form
\begin{eqnarray} \label{eq:kahler-metric_2}
K_{i \bar{j}} 
 =  \frac{3}{4 u_1^2 \left(1 - \epsilon^3 \right)^2} 
\Biggl( \begin{array}{cc} 1 + \frac{\epsilon^3}{2} & - \frac{3 \epsilon}{2}
\\ - \frac{3 \epsilon}{2} & \epsilon + \frac{1}{2 \epsilon} \end{array} \Biggr) \; ,
\end{eqnarray}
retaining only the leading powers of $\epsilon$, we can find the eigenvalues and hence see that
\begin{eqnarray}
f_{\rm light}^2 &=& \frac{3}{4 u_1^2} \; ,\\
f_{\rm heavy}^2 &=& \frac{f_{\rm light}^2}{2 \epsilon} = \frac{\epsilon^3}{2}\frac{3}{4 u_2^2}  \; ,
\end{eqnarray}
where, the $f_{\rm heavy}$ does not become too large.
At leading order, the matrix of change of basis is
\begin{equation}
P \approx \Biggl( \begin{array}{cc} 1 - {\cal O}(\epsilon^4)  & -{\cal O}(\epsilon^4)
\\ 3 \epsilon^2 + {\cal O}(\epsilon^3)  & 1 + {\cal O}(\epsilon^4)  \end{array} \Biggr) \; ,
\end{equation}
where, we follow the convention that the first column of $P$ is the eigenvector corresponding to smaller eigenvalue. Let us suppose that when we try to retain the two-axion limit, the search direction we explore is the intersection of the plane of perturbatively unfixed axions and $\sigma-{\tilde \nu}_1$ plane, this makes sense since this is equivalent to ${\tilde \nu}_2 = 0$.
Using the above form of the $P$ matrix in Eq (\ref{eq:plane}), it is easy to see that, in the limit $\epsilon \rightarrow 0$, the line which is common to this plane (in the space of canonically normalised scalar fields) and the plane ${\tilde \nu}_2 = 0$ has along the vector $(1,-1/\sqrt{3},0)$. Since this is the direction along which $\psi$ is defined in the two-axion limit, this indicates that we have recovered the flux independence of the slope of this line (mentioned in \textsection \ref{sec:flux-ind}) in the two-axion limit.
Moreover, if we keep the leading order terms in $\epsilon$ and follow the procedure described in \textsection \ref{sec:my_formalism}, we can see that 
\begin{equation}
 f_\psi^{u_1} = 2 f_{\tilde \nu_1} + {\cal O}(\epsilon^3) \; .
\end{equation}
This result was also mentioned in \cite{Palti:2015xra}. 
Now, having recovered the results in the two-axion case, let us apply the ideas presented in \textsection \ref{sec:my_formalism}.

\begin{figure*}
  \includegraphics[width = 0.9\textwidth]{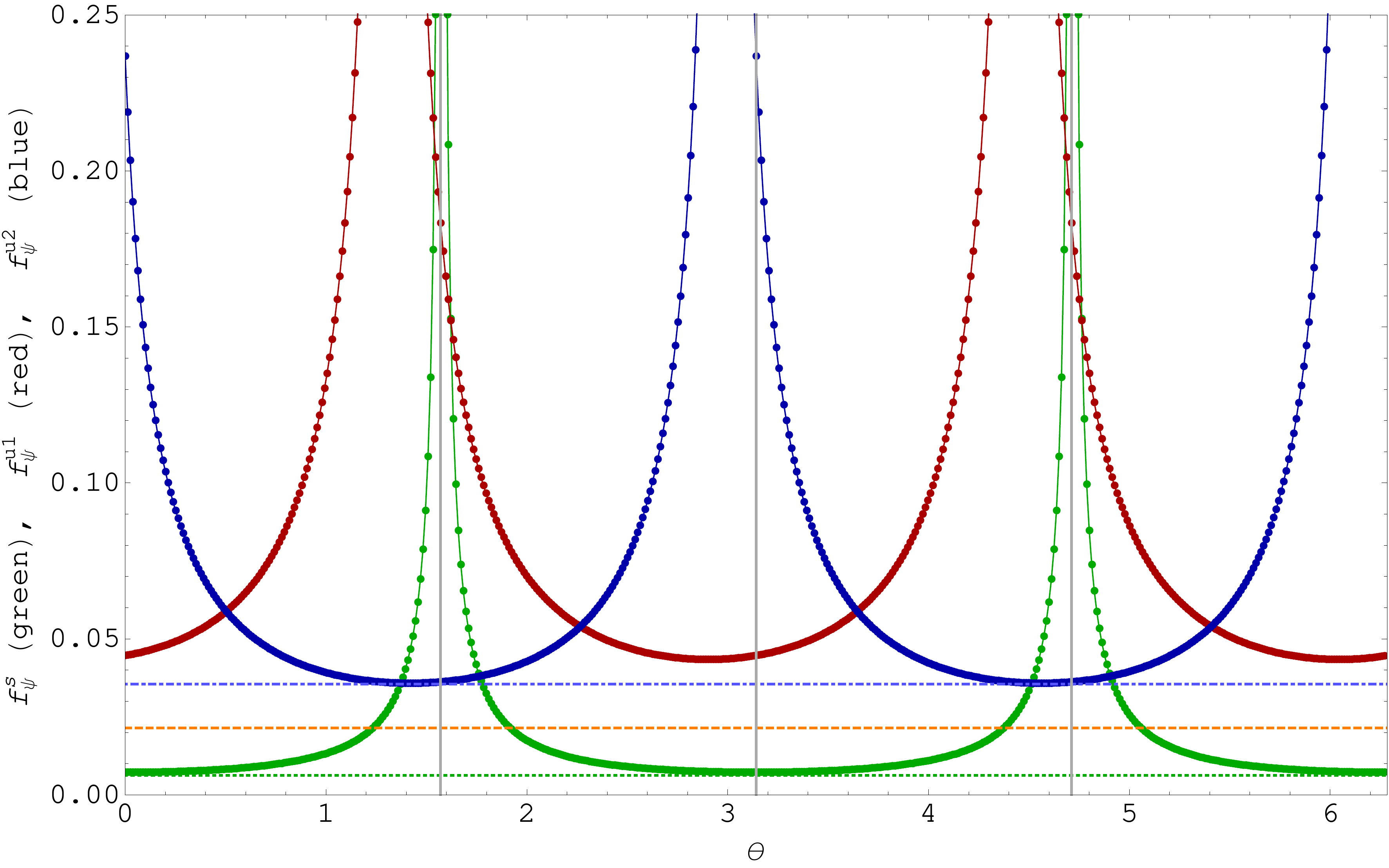}
  \caption
 {For the fixed choice of fluxes mentioned in the text, we can explore different directions in field space by numerically implementing the formalism described in \textsection \ref{sec:my_formalism}. As we vary the angle $\theta$, we get enhancement in all three effective decay constants $f_\psi^{s}, f_\psi^{u_1}$ and $f_\psi^{u_2}$.
 The dotted green horizontal line is the value $f_\sigma$, the dashed orange horizontal line is $f_{\tilde \nu_1}$ while the dot-dashed blue horizontal line is the value of $f_{\tilde \nu_2}$. Notice that when $\theta$ is zero, we get roughly recover the two-axion limit. The vertical lines correspond to $\theta$ being $\pi/2$, $\pi$ and $3\pi/2$.
 }
  \label{angular}
\end{figure*}

To this end, we begin to explore other directions in the three axion field space. In generating fig (\ref{angular}), we have fixed the fluxes to the following values $q_1 = 60, q_2 = 12, h_0 = 10, f_0 = 10$ and the volume ${\cal V} \approx 200$. This gives $\epsilon = 0.2$ and the initial direction of exploration (the intersection of the plane of perturbatively unfixed axions and $\sigma-{\tilde \nu}_1$ plane) makes an angle of -29.25 deg w.r.t. $\sigma$ direction, which is pretty close to the angle obtained in \textsection \ref{sec:flux-ind}. 

As we explore other directions in the plane by changing the angle $\theta$ (see discussion below eq (\ref{eq:initial_search_dir})), we experience enhancement of the decay constants. 
Furthermore, the case $\theta = 0$ gives results approximately in agreement with the two-axion case. 
In fig (\ref{angular}) the minimum of the green curve (the variation of $f_\psi^{s}$ against $\theta$) lies very close to the dotted green horizontal line (which specifies $f_\sigma$). But the the minimum of the red curve (the variation of $f_\psi^{s}$ against $\theta$) is far above the dashed orange horizontal line (which specifies $f_{\tilde \nu_1}$). This is a manifestation of eq (\ref{eq:2-axion-eff-decay}) and Eq (\ref{eq:2-axion-eff-decay_2}). In particular, in fig (\ref{angular}), at $\theta = 0$, the extreme left region, $ f_\psi^{u_1} \approx 2 f_{\tilde \nu_1}$. Similarly, is is easy to see from fig (\ref{angular}) that, as mentioned in \textsection \ref{sec:remarks}, as we vary $\theta$, the effective decay constants blow up and this happens for $\theta = \pi/2$ for $f_\psi^{s}$.


While it was useful to recover the two-axion model in an appropriate limit of the three-axion model, one must understand that one could vary fluxes such that $\epsilon$ is no longer small. For every choice of fluxes, we could vary $\theta$ to look for directions to enhance the effective decay constants. The results presented here suggest that the decay constants can be enhanced this way. In particular, no matter what choice of fluxes on begins with, one could always vary $\theta$ and find directions in field space along which the potential seems quite flat.



\subsection{Failure of the approach} \label{sec:failure}

The results presented till now basically imply that there are some radial directions in the perturbatively flat plane in field space in which the potential appears to be a cosine with arbitrarily large effective decay constant. For certain directions, this effective decay constant could even be infinite! The question is: is this the correct way to look for large effective decay constants?
We are of course free to explore any direction in field space, but there is no reason that the field would actually evolve in the direction which we choose to explore. Will the field roll in the direction along which the effective decay constant is large? 

\begin{figure}
  \includegraphics[width = 0.4\textwidth]{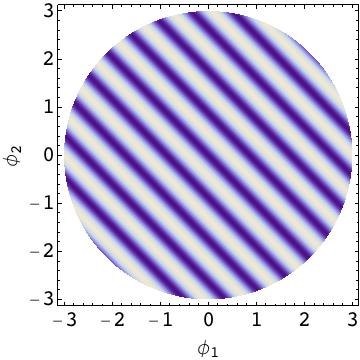}
  \caption
 {This is the density plot of the potential in Eq (\ref{eq:Lag_low_3_psi}) with $\psi$ being the radial coordinate and $\theta$ being the angular coordinate. The radial coordinate $\psi$ varies from 0 to 3 while the angular coordinate goes from 0 to 2$\pi$. Here, $\phi_1 = \psi \cos \theta$ and $\phi_2 = \psi \sin \theta$.
 Notice the flat directions in the field space, these flat directions would lead to the effective decay constant being interpreted as infinite in certain directions.
 }
  \label{contour_fail}
\end{figure}

In the language of plane polar coordinates introduced in the perturbatively flat plane, the scalar field $\psi$ is the radial coordinate in this plane while $\theta$ will be the angular coordinate and the potential for $\psi$ in Eq (\ref{eq:Lag_low_3_psi}) depends on the choice of $\theta$.
In order to better understand what's happening, we could find the potential in Eq (\ref{eq:Lag_low_3_psi}) for various values of $\theta$ and then plot the potential for all these values of $\theta$.  
When we do that, we obtain the contour plot of the potential as shown in fig (\ref{contour_fail}).
It now becomes clear the Lagrangian in Eq (\ref{eq:Lag_low_3}) happens to be such that one direction in the plane is still a flat direction. This is the reason because of which we get the effective decay constant which diverges for certain choices of $\theta$ and this is what explains the results shown in fig (\ref{angular}). Thus, the apparent enhancement in effective decay constant is just an artefact of the fact that the Lagrangian of Eq (\ref{eq:Lag_low_3}) does not lead to stabilization of all the moduli.
Finally, notice that in fig (\ref{contour_fail}), we have introduced the fields $\phi_1$ and $\phi_2$, which are Cartesian coordinates in the perturbatively flat plane in field space. 

Thus, the approach described in the last subsection does not lead to any actual enhancement of the effective decay constant. But, we can learn the following important lessons from this:
\begin{enumerate}
 \item we should only pay attention to the behaviour of the scalar potential in the directions in which the field could actually roll,
 \item we should avoid situations in which there are massless directions in field space, and,
 \item since it is easier to interpret our results if we work with fields which act as Cartesian coordinates in the perturbatively flat plane, we must work with such fields.
\end{enumerate}

\section{A more careful attempt to obtain large effective decay constant} \label{sec:careful}

In this section, we shall try another, much more careful approach to obtain large effective decay constant which will not suffer from the problems of the approach explained in the last section. As we shall see, even this approach will fail to give a super-Planckian effective decay constant.

\subsubsection{Evolution in field space}

We are interested in the behaviour of fields near AdS vacua. 
For a field theory with negative vacuum energy, the spacetime evolves as a spatially open FRW Universe (which expands and then re-collapses). If space is homogeneously filled with the scalar fields which roll down their potentials (which is negative), the fields and spacetime evolve in accordance with 
\begin{eqnarray}
\ddot{\phi}_i + 3 H \dot{\phi}_i + \left( \frac{\partial V}{\partial \phi_i} \right) &=& 0 \; , \\
H^2 - \frac{1}{a^2} - \frac{8 \pi G}{3} \left[ \frac{\gamma_{n m} \dot{\phi}_n \dot{\phi}_m}{2} + V(\phi) \right] &=& 0\; ,
\end{eqnarray}
where, we have set $K= -1$. In the following, we shall ignore the evolution of spacetime and hence, for us,
\begin{equation}
\ddot{\phi}_i = - \left( \frac{\partial V}{\partial \phi_i} \right) \; ,
\end{equation}
i.e. the acceleration of the field in field space will point in the direction opposite to that of the gradient vector.
At a local minimum in field space, the gradient vanishes so the acceleration of the field at a point near the local minimum would be given by 
\begin{equation}
\ddot{\phi}_i |_{nearby} = - \left( \frac{\partial V}{\partial \phi_i} \right)_{nearby}  \approx \left( \frac{\partial^2 V}{\partial \phi_i \partial \phi_j} \right)_0 \Delta \phi_j \; ,
\end{equation}
so that if the direction happens to be an eigenvector of the Hessian matrix of the field, the field accelerates in the same direction as the displacement. Furthermore, if the initial velocity of the field is zero, it evolves in the same direction as acceleration.
For potential given by Eq (\ref{eq:pot_KNP}), near the origin, if one finds the eigenvectors of the Hessian matrix, when $|f_1 g_2 - f_2 g_1|$ becomes small, these eigenvectors coincide with $\psi_1$ and $\psi_2$ directions.

Thus, there will now be no freedom to explore the various directions in field space. When all moduli are stabilized, the Hessian matrix will have two positive eigenvalues, we must look at the behaviour of the potential in the direction of the eigenvector corresponding to smaller eigenvalue.

\subsubsection{Realising KNP mechanism} \label{sec:KNP}

From the Kahler potential, Eq (\ref{eq:kahler_pot}), with ${\cal V}'$ given by Eq (\ref{eq:cmplx_vol}), one can easily find the Kahler metric, which in this case will be given by
\begin{eqnarray} \label{eq:kahler-metric}
K_{i \bar{j}} 
 = \alpha  \left(\begin{array}{ccc} \frac{1}{4s^2 \alpha} & 0 & 0 \\
0 & 6u_1 + \frac{3 u_2^{3/2}}{u_1^{1/2}}  & -9 \sqrt{u_1 u_2}\\ 
0 & -9 \sqrt{u_1 u_2} & 6u_2 + \frac{3 u_1^{3/2}}{u_2^{1/2}} 
\end{array}\right) \; ,
\end{eqnarray}
where $\alpha = \frac{1}{8} \left( u_1^{3/2} - u_2^{3/2} \right)^{-2}$. Notice that the metric still depends on the vev of the moduli $s$, $u_1$ and $u_2$.
To aid the discussion, let us denote the set of fields $(\sigma, \nu_1, \nu_2)$ by $(\chi_1, \chi_2, \chi_3)$. Then, we could assume that appropriate D-brane instantons have been chosen such that the Lagrangian of the low energy effective theory is of the following form
\begin{eqnarray} \label{eq:Lag_low_1}
{\cal L} = &-& \frac{1}{2} K^{ij} (\partial \chi_i) (\partial \chi_j) 
- \Bigl[ V_0 + A' e^{-s} (1-\cos \chi_1)
\nonumber \\
&+& B' e^{-u_1} (1- \cos {\chi}_2) + C' e^{-u_2} (1- \cos {\chi}_3) \Bigr] \; ,
\end{eqnarray}
where the Kahler metric is not diagonal. The above choice happens to be such that there are no massless directions in field space anymore.
The factor of half in front of the kinetic term arises from the fact that the complex scalars in Eq (\ref{eq:kin}) are decomposed as $\phi_i = \frac{1}{\sqrt 2} (\phi^R_i + i \phi^I_i)$ etc.

Since the metric would still be a real, symmetric matrix, it can be diagonalised by an orthogonal transformation in the field space. 
In this choice of basis fields, the Kahler metric (which in this case would be a real and symmetric matrix) is not diagonal, it then makes sense to make an orthogonal transformation in field space to diagonalise it. Let the new basis fields be $\xi_l$, then, 
\begin{equation} \label{eq:P}
\chi_i = P_{i l} \xi_l \; ,
\end{equation}
and hence the Lagrangian in Eq (\ref{eq:Lag_low_1}) becomes 
\begin{eqnarray} \label{eq:Lag_tmp_1}
{\cal L} = &-& \frac{1}{2} (\partial \xi_l) P^T_{l i} K^{i j} P_{j k} (\partial \xi_k) \nonumber \\
&-& \Bigl[ V_0 + A' e^{-s} \left\{1-\cos ( P_{1 l} \xi_l )  \right\} \nonumber \\
&+& B' e^{-u_1} \{1- \cos ( P_{2 l} \xi_l ) \} \nonumber \\
&+& C' e^{-u_2} \{ 1- \cos ( P_{3 l} \xi_l ) \} \Bigr] \; .
\end{eqnarray}
It is well known that if the matrix $P$ is chosen such that its $k^{\rm th}$ column is $k^{\rm th}$ eigenvector of $K$, then,
$P^T K P$ will be a diagonal matrix $D$. Let 
\begin{equation}
 D = {\rm diag} ( d_1^2, d_2^2, d_3^2 ) \; ,
\end{equation}
then, Eq (\ref{eq:Lag_tmp_1}) takes the form 
\begin{eqnarray} \label{eq:Lag_tmp_2}
{\cal L} = &-& \frac{1}{2} d_i^2 (\partial \xi_i)^2 - V(\xi_1, \xi_2, \xi_3) \; .
\end{eqnarray}
Next one could redefine the fields such that the Kahler metric becomes an identity matrix: this requires a scaling transformation in the field space. 
We could now make a scaling transformation of the form 
\begin{equation}
\psi_i = d_i \xi_i ~~~~({\rm no ~sum ~over ~}i) \; ,
\end{equation}
and hence Eq (\ref{eq:Lag_tmp_1}) becomes
\begin{eqnarray} \label{eq:Lag_tmp_2}
{\cal L} = &-& \frac{1}{2} (\partial \psi_l)^2  \nonumber \\
&-& \Bigl[ V_0 + A' e^{-s} \left\{1-\cos \left( \frac{P_{1 l} \psi_l}{d_l} \right)  \right\} \nonumber \\
&+& B' e^{-u_1} \left\{1- \cos \left( \frac{P_{2 l} \psi_l}{d_l} \right) \right\} \nonumber \\
&+& C' e^{-u_2} \left\{ 1- \cos \left( \frac{P_{3 l} \psi_l}{d_l} \right) \right\} \Bigr] \; ,
\end{eqnarray}
where, the sum over $l$ is implied. Note that the field space metric for $\psi_l$ fields is a Kronecker delta. 

One could find the equation of the perturbatively flat plane in terms of the new fields. The constraint of always being in this plane can be used to eliminate one of the field using the equation of the plane. One can then introduce a ``Cartesian-coordinate" system in this plane and write the Lagrangian in terms of the the fields.
This Lagrangian depends on the three fields $\psi_l$, but, since there is a constraint of being stuck in the plane described by Eq (\ref{eq:condition}), there are really only two independent fields.
Since we are stuck in the plane Eq (\ref{eq:condition}), it is sensible to introduce a Cartesian coordinate system in this plane and express the above Lagrangian, Eq (\ref{eq:Lag_tmp_2}) in terms of these fields. To this end, we note that Eq (\ref{eq:condition}) can be written in terms of $\psi_i$ as
\begin{equation} \label{eq:plane_new}
A_1 \psi_1 + A_2 \psi_2 + A_3 \psi_3 = 0 \; ,
\end{equation}
with
\begin{eqnarray}
A_1 &=& \frac{(h_0 P_{1 1} + q_1 P_{2 1} + q_2 P_{31})}{d_1} \; , \\
A_2 &=& \frac{(h_0 P_{1 2} + q_1 P_{2 2} + q_2 P_{32})}{d_2} \; , \\
A_3 &=& \frac{(h_0 P_{1 3} + q_1 P_{2 3} + q_2 P_{33})}{d_3} \; .
\end{eqnarray}
This is the equation of the plane described by Eq (\ref{eq:condition}), which we shall denote as $P$, in terms of the new fields. Note that the Eq (\ref{eq:plane_new}) implies that the vector 
\begin{equation}
{\vec N} = (A_1, A_2, A_3) \; ,
\end{equation}
must be normal to the plane $P$.
Now consider the intersection of the plane $P$ with the plane $\psi_3 = 0$, this gives a line, a vector along this line would be such that $A_1 \psi_1 + A_2 \psi_2 = 0$, so, it is of the form
\begin{equation}
{\vec V} = \left(1, - \frac{A_1}{A_2}, 0 \right) \; ,
\end{equation}
unlike the vector $\vec N$, the vector $\vec V$ lies in the plane $P$. In order to have an orthonormal basis, we need another vector which lies in the plane $P$ which is perpendicular to $\vec V$. We can consider
\begin{equation}
{\vec W} = {\vec N} \times {\vec V} \; ,
\end{equation}
then, 
\begin{equation}
 {\vec W} = \left( \frac{A_1 A_3}{A_2}, A_3, - \frac{A_1^2 + A_2^2}{A_2} \right) \; ,
\end{equation}
it is easy to verify that the components of this vector satisfy Eq (\ref{eq:plane_new}).
From the expressions for $\vec N$, $\vec V$ and $\vec W$, one can find the unit vectors $\hat N$, $\hat V$ and $\hat W$.
Let us write ${\hat V} = (V_1, V_2, V_3)$, ${\hat W} = (W_1, W_2, W_3)$ and ${\hat N} = (N_1, N_2, N_3)$, notice that $V_3 = 0$.

Now consider an arbitrary vector $\vec U$ in the field space,
\begin{equation}
{\vec U} = \psi_1 {\hat e}_1 + \psi_2 {\hat e}_2 + \psi_3 {\hat e}_3 \; ,
\end{equation}
if we use $({\hat V}, {\hat W}, {\hat N})$ as the new basis, then, the same vector could be written as
\begin{equation}
{\vec U} = \phi_1 {\hat V} + \phi_2 {\hat W} + \phi_3 {\hat N} \; .
\end{equation}
One can use the above two expressions for $\psi_i$ to $\psi_j$ etc.If the vector $\vec U$ lies in the plane $P$, then,
$\phi_3 = 0$ and we get
\begin{eqnarray}
\psi_1 &=& \phi_1 V_1 + \phi_2 W_1 \; , \\
\psi_2 &=& \phi_1 V_2 + \phi_2 W_2 \; , \\
\psi_3 &=& \phi_2 W_3 \; .
\end{eqnarray}
We could now use these expressions in Eq (\ref{eq:Lag_tmp_2}) and get the following

\begin{eqnarray} \label{eq:Lag_KNP_realised}
{\cal L} = &-& \frac{1}{2} (\partial \phi_1)^2 -\frac{1}{2} (\partial {\phi}_2)^2 \nonumber \\
&-& \Bigl[ V_0 + A' e^{-s} \left( 1- \cos \left[ \frac{\phi_1}{f_1} + \frac{\phi_2}{g_1} \right] \right) \nonumber \\
&+& B' e^{-u_1} \left( 1 - \cos \left[ \frac{\phi_1}{f_2} + \frac{\phi_2}{g_2} \right] \right) \nonumber \\
&+& C' e^{-u_2} \left( 1- \cos \left[ \frac{\phi_1}{f_3} + \frac{\phi_2}{g_3} \right] \right) \Bigr] \; ,
\end{eqnarray}
where,
\begin{eqnarray} \label{eq:decay_const}
\frac{1}{f_3} = \frac{P_{11} V_1}{d_1} + \frac{P_{12} V_2}{d_2} \; , \frac{1}{g_3} = \frac{P_{11} W_1}{d_1} + \frac{P_{12} W_2}{d_2}+ \frac{P_{13} W_3}{d_3} \; , \nonumber \\ 
\frac{1}{f_2} = \frac{P_{21} V_1}{d_1} + \frac{P_{22} V_2}{d_2} \; , \frac{1}{g_2} = \frac{P_{21} W_1}{d_1} + \frac{P_{22} W_2}{d_2}+ \frac{P_{23} W_3}{d_3} \; , \nonumber \\ 
\frac{1}{f_1} = \frac{P_{31} V_1}{d_1} + \frac{P_{32} V_2}{d_2} \; , \frac{1}{g_1} = \frac{P_{31} W_1}{d_1} + \frac{P_{32} W_2}{d_2}+ \frac{P_{33} W_3}{d_3} \; .  
\end{eqnarray}

From Eq (\ref{eq:Lag_KNP_realised}), it is clear that this is a successful realisation of KNP mechanism.

\subsubsection{Numerical analysis}

\begin{figure*}
  \includegraphics[width = 0.5\textwidth]{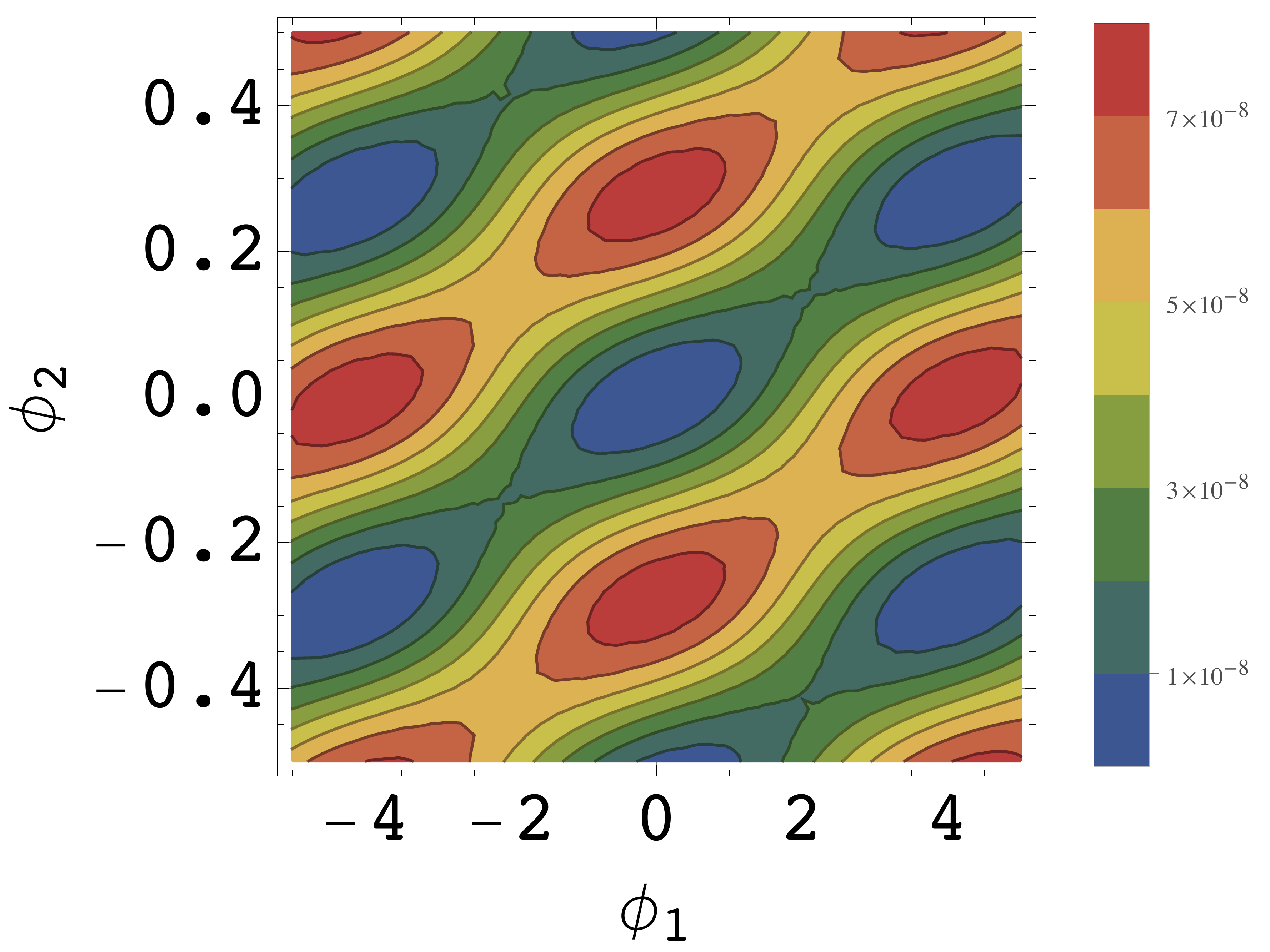}
  \caption
 {The contour plot of the scalar potential in Eq (\ref{eq:Lag_KNP_realised}) in $(\phi_1,\phi_2)$ field space for the flux choices $q_1 = 34$, $q_2 = 33$ (as well as $f_0 = 10$, $h_0 = 10$ and the volume ${\cal V} = 100$). Note the difference in the scale of the horizontal axis and vertical axis. The corresponding $f_{\rm eff} = 0.663$, thus the distance between two local minima of the potential is approximately $2 \pi f_{\rm eff} \approx 4.16 M_{\rm Pl}$.
 }
  \label{contour}
\end{figure*}

From Eq (\ref{eq:cmplx_vol}), it is clear that $u_1$ must be greater than $u_2$, Eq (\ref{eq:u1}) and Eq (\ref{eq:u2}) then suggest that this can be ensured if $q_1 > q_2$. We fix the fluxes $f_0$ and $h_0$ and fix the volume ${\cal V}$. This fixes the value of the modulus $s$. We now wish to change the fluxes $q_1$ and $q_2$ (with $q_1 > q_2$) and for each choice of these fluxes, we can find the values of the fields $u_1$ and $u_2$. 

If we now work in the regime in which $s \gg u_1 > u_2$ so that the first cosine in Eq (\ref{eq:Lag_KNP_realised}) gives negligible contribution and can be completely ignore. In this regime, Eq (\ref{eq:Lag_KNP_realised}) becomes identical to Eq (\ref{eq:pot_KNP}) where the decay constants are completely determined by fluxes.
To achieve this, we only admit those choices of fluxes which ensure that $s \gg u_1 > u_2$. 
Furthermore, we would enforce the constraint that the ratio $e^{-u_2}/e^{-u_1}$ lies between 1 and 10. Otherwise, the amplitudes of the two cosines in Eq (\ref{eq:pot_KNP}) would be too different even without the alignment mechanism.

One can now determine the Kahler metric Eq (\ref{eq:kahler-metric}) and hence its eigenvalues, eigenvectors and the elements of the matrix $P$ in Eq (\ref{eq:P}). In trustworthy regimes, we expect the eigenvalues of the Kahler metric to be less than one in Planck units, because they are closely related to Volumes of cycles which better be large in string units.
Thus, we also enforce the requirement that the quantities $d_1$, $d_2$ and $d_3$ are all smaller than 1.

Keeping all of these observations in mind, it appears that, since we have so many fluxes available to adjust, we should be able to get $f_{\rm eff}$ to be as large as desired. 
Thus, we fix the fluxes $f_0 = 10$, $h_0 = 10$ and the volume ${\cal V} = 100$, this then fixes the modulus $s = 40$.
We then vary $q_1$ from 26 to 225 while, for each of these values of $q_1$, we let $q_2$ vary from 25 to $q_1 - 1$ (to ensure that $q_1 > q_2$).
This leads to 20100 combinations of values of fluxes, of course, not all these combinations would lead to the satisfaction of the constraints outlined above. Among these, there are only 3100 flux combinations which lead to the satisfaction of all the constraints mentioned. 

Next, for each of these choices of the fluxes $q_1$ and $q_2$ (with $q_1 > q_2$), one could use the formalism presented in the last subsection to determine the ``decay constants" $f_1, f_2, f_3$ as well as $g_1, g_2, g_3$. Since these ``decay constants"are derived quantities with no geometrical interpretation, we do not enforce the constraints that they be small or large. We shall find that these derived ``decay constants" could also sometimes be negative.
Note that, if $P_{11}$ and $P_{12}$ both happen to be zero, then $f_1$ would be infinity. Thus, since the matrix elements of the matrix $P$ turn up in the expressions of the decay constants, the values of these derived decay constants may appear to become unreasonably large for some circumstances.
Having determined these derived decay constants, we could find whether $|f_1g_2-f_2g_1|$ is sufficiently small. If that happens to be the case, the value of $f_{\rm eff}$ can be found from Eq [\ref{eq:feff}]. On the other hand, when $|f_1g_2-f_2g_1|$ is not too small, there is no hierarchy between the masses of $\psi_1$ and $\psi_2$ and hence no sufficiently flat direction in the field space.

We could now impose the requirement that we'd only consider those values of fluxes which lead to the value of $|f_1g_2-f_2g_1|$ to be less than, 0.25 (to ensure sufficient alignment), then, we find that there are only 68 combinations of fluxes which satisfy this constraint. Among these, the largest value of effective decay constant (i.e. Eq (\ref{eq:feff})) obtained is $f_{\rm eff} = 0.663$, corresponding to $q_1 = 34$ and $q_2 = 33$ and the corresponding alignment angle $\Delta \theta$ in Eq (\ref{eq:delta_theta}) is $\Delta \theta = 7.63$ degrees (i.e. roughly 0.13 radian).
The behaviour of the scalar potential near the origin in $(\phi_1,\phi_2)$ space for this choice of fluxes can be seen in fig (\ref{contour}).

Thus, provided we satisfy all the constraints mentioned above, even though we have the freedom to adjust fluxes to change the values of the individual decay constants which turn up in Eq (\ref{eq:Lag_KNP_realised}), we can not make the effective decay constant to be super-Planckian and we can not decrease the alignment angle to an arbitrarily small value.

\section{Discussion} \label{sec:discussion}

It has been observed for a very long time \cite{Banks:2003sx} that obtaining a large axionic decay constant i.e. $f \gg M_p$ for string theory axions always takes us out of the trustworthy regime of theoretical control. 
All of this is closely related to a number of issues studied in the recent literature e.g. the various versions of swampland distance conjecture and axionic weak gravity conjecture \cite{Klaewer:2016kiy,Font:2019cxq,Grimm:2019wtx}
as well as issues of the trajectory being followed in field space \cite{Landete:2018kqf}.
Keeping this literature in mind, in the present work, we revisited the question of possible enhancement of the effective decay constant of axions. The best interpretation of the results we have obtained is that we have found more evidence that in trustworthy regimes of string theory, the effective decay constant can not be super-Planckian. 

One must note the various caveats associated with the formalism used in this work: firstly, in these constructions, the axionic field space is not compact e.g. in two-axion case, the the potential in direction orthogonal to the perturbatively flat direction is not periodic. 
Secondly, it seems that the conclusions are too much dependent on the choice of the specific Euclidean D-brane instantons.
Also, note that the type IIA flux vacua we have been dealing with are AdS. Thus, if one intends to make any statement about cosmic inflation, one must find what uplifting will do to the various effects described here. Thus, one needs to understand uplifting mechanisms better before one can make any statements about large field inflation based on the work of this paper.
Finally,, one must mention some of the concerns expressed in the literature about the validity of the solutions used in this paper \cite{Acharya:2006ne,Banks:2006hg,Saracco:2012wc,McOrist:2012yc,Blaback:2010sj,Gautason:2015tig,Andriot:2018ept,Junghans:2018gdb,Banlaki:2018ayh,Cordova:2018dbb,Escobar:2018rna}.
Since in such constructions one works with massive type IIA supergravity (supplemented with orientifold 6-plane sources), the corresponding Romans mass parameter does not dilute in the large volume limit, this has inspired doubts about the validity of these constructions. 
In addition to the problems with expanding around a non-solution like a Calabi-Yau metric or the concerns about defining O6-planes \cite{Acharya:2006ne}, one also finds that the solutions of \cite{DeWolfe:2005uu} do not solve the massive IIA supergravity equations of motion even approximately \cite{McOrist:2012yc} (as required for large volume, weakly-coupled backgrounds). We must also mention that we have ignored the open string moduli arising from e.g. brane positions as well as blow-up moduli or twisted moduli, we have not attempted to enforce tadpole cancellation conditions and we have not taken into account the backreaction caused by Kahler moduli as pointed out in \cite{Hebecker:2018fln} very recently.

With all of these issues in mind,  \textsection \ref{sec:3axion}, we tried to explore the possibility of enhancement of axion decay constant from the point of view of the statement of axionic Weak Gravity Conjecture. 
Thus, we tried to find directions in axion field space in which the scalar potential is a cosine with a large period. This requires two things to happen (a) the potential along the direction of interest must be a cosine with effective decay constant being due to only one of the axions and being large, and, (b) the vev of the saxion corresponding to the rest of axions must be so large that their contributions to the scalar potential must be negligible. We found that just these requirements can always be fulfilled for any fixed choice of fluxes. But, as we explained in \textsection \ref{sec:failure}, we realised that we should be careful to ensure that the field should actually roll along the direction so found and to ensure that no moduli are left unstabilised.

In order to cure these problems, we came up with a completely explicit realisation of the KNP alignment mechanism in the context of type IIA string theory (see Eq (\ref{eq:Lag_KNP_realised})) which proves to be a very convenient way of obtaining an effective axion decay constant in a well-controlled regimes of string theory. 
The decay constants in this case can be explicitly found for any choice of fluxes (see e.g. the discussion around Eq (\ref{eq:Lag_KNP_realised})).
We have thus presented a formalism which can be used to find the alignment angle $\Delta \theta$ (defined by Eq (\ref{eq:delta_theta})) as well as the effective decay constant $f_{\rm eff}$ (defined by (i.e. Eq (\ref{eq:feff})) from the fluxes.
It might appear that, since we have so many fluxes available to adjust, we should be able to get $f_{\rm eff}$ to be as large as desired. But when we numerically evaluated  quantities for a very large combination of fluxes and imposed some sensible conditions on the valid answers, $\Delta \theta$ and $f_{\rm eff}$, it was found (as is suggested by a lot of previous literature) that unlike any randomly chosen field theory, in a low energy effective field theory arising from string theory, somehow, there seems to be a sub-Planckian upper limit on the effective decay constant and a lower limit on the alignment angle.
This provides yet another example of the fact that the famous KNP alignment mechanism does not quite work in explicit examples in string theory.

 \noindent{\bf Acknowledgements}
The author would like to thank Mansi Dhuria (IITRAM, India), Ashwin Pande (Ahmedabad University, India), Anshuman Maharana (HRI, India), Ashoke Sen (HRI, India), Timm Wrase (TU Wien, Austria), Fernando Quevedo (ICTP, Italy), Eran Palti (MPP, Germany) and Chethan Krishnan (IISc, India) for clarifying a number of issues related to string compactifications. 
This research was supported in part by the International Centre for Theoretical Sciences (ICTS) during a visit for participating in the program - Kavli Asian Winter School (KAWS) on Strings, Particles and Cosmology 2018 (Code: ICTS/Prog-KAWS2018/01). 
The author would also thank the organisers of School and Workshop on ``String Field Theory and String Phenomenology" held at Harish Chandra Research Institute (HRI), Allahabad, India. 
 \\ \\

\appendix*

\end{document}